\documentclass[twocolumn,tighten,times]{aastex63}
\usepackage{graphicx}
\usepackage{amsmath}
\usepackage{natbib}
\usepackage{amssymb}
\usepackage{sidecap}

\shorttitle{Toy Model for Fast Radio Bursts}
\shortauthors{Metzger et al.}

\begin{document}

\newcommand{\be}{\begin{equation}}
\newcommand{\ee}{\end{equation}}

\title{A Toy Model for the Time-Frequency Structure of Fast Radio Bursts: Implications for the CHIME Burst Dichotomy}

\author[0000-0002-4670-7509]{Brian D. Metzger}
\affil{Department of Physics and Columbia Astrophysics Laboratory, Columbia University, Pupin Hall, New York, NY 10027, USA}
\affil{Center for Computational Astrophysics, Flatiron Institute, 162 5th Ave, New York, NY 10010, USA} 

\author[0000-0002-5519-9550]{Navin Sridhar}
\affiliation{Department of Astronomy and Columbia Astrophysics Laboratory, Columbia University, New York, New York 10027, USA}

\author[0000-0001-8405-2649]{Ben Margalit}
\altaffiliation{NASA Einstein Fellow}
\affiliation{Astronomy Department and Theoretical Astrophysics Center, University of California, Berkeley, Berkeley, CA 94720, USA}

\author[0000-0001-7833-1043]{Paz Beniamini}
\affiliation{Astrophysics Research Center of the Open University (ARCO) and Department of Natural Sciences, The Open University of Israel, P.O Box 808, Ra’anana 43537, Israel}
\affiliation{Theoretical Astrophysics, Walter Burke Institute for Theoretical Physics, Mail Code 350-17, Caltech, Pasadena, CA 91125, USA}

\author[0000-0002-1227-2754]{Lorenzo Sironi}
\affiliation{Department of Astronomy and Columbia Astrophysics Laboratory, Columbia University, New York, New York 10027, USA}

\begin{abstract}
We introduce a toy model for the time-frequency structure of fast radio bursts (FRB), in which the observed emission is produced as a narrowly-peaked intrinsic spectral energy distribution sweeps down in frequency across the instrumental bandpass as a power-law in time.  Though originally motivated by emission models which invoke a relativistic shock, the model could in principle apply to a wider range of emission scenarios.  We quantify the burst's detectability using the frequency bandwidth over which most of its signal-to-noise ratio (SNR) is accumulated.  We demonstrate that by varying just a single parameter of the toy model$-$the power-law index $\beta$ of the frequency drift rate$-$one can transform a long (and hence preferentially time-resolved) burst with a narrow time-integrated spectrum into a shorter burst with a broad power-law time-integrated spectrum.  We suggest that burst-to-burst diversity in the value of $\beta$ could generate the dichotomy between burst duration and frequency-width recently found by CHIME.  In shock models, the value of $\beta$ is related to the radial density profile of external medium, which in light of the preferentially longer duration of bursts from repeating sources may point to diversity in the external environments surrounding repeating versus one-off FRB sources.
\end{abstract}


\section{Introduction}

CHIME/FRB reported in their first catalog paper (\citealt{CHIME+21_catalog,Pleunis+21}) an apparent dichotomy in the observed properties of fast radio bursts (FRB), which supports earlier suggestions based on smaller sample sizes (e.g.,~\citealt{Scholz+16,CHIME+19repeaters,Hashimoto+20}).  In particular, they find that FRBs with longer durations are observed to (1) possess narrower time-integrated spectra which often peak in the middle of the instrumental band-pass; (2) preferentially be associated with FRB sources thus far observed to repeat.  By contrast, shorter bursts possess power law-like spectra covering most or all of the instrumental band-pass and are more frequently associated with non-repeating one-off bursts (i.e., sources not yet observed to repeat).  Furthermore, while temporally resolved bursts from repeaters exhibit narrow $\lesssim 100$ MHz spectral structures (``sub-bursts'') that often drift downwards in frequency with time (e.g.,~\citealt{Michilli+18,Gajjar_18,Hessels+19,Josephy+19,CHIME+19repeaters}), this behavior is seen less frequently in the non-repeating sources.  

On the other hand, the distribution of dispersion measures (DM) of the repeating and non-repeating populations are broadly similar \citep{CHIME+21_catalog}, consistent with a similar range of source distances.  Although instrumental effects can in some cases distort the observed burst properties (e.g., broad-band bursts that occur in the instrumental side-lobes may exhibit artificially narrow spectra), the dichotomy observed by CHIME was argued to be robust to these contaminants \citep{Pleunis+21}.  If intrinsic, this difference in repeating and non-repeating classes could point to fundamental diversity in the FRB central engines and/or emission mechanisms.

A large number of FRB models have been proposed in the literature (e.g., \citealt{Cordes&Chatterjee19,Platts+19,Petroff+19,Petroff+21} for reviews), but at a basic level many of these models attribute the radio emission to a sudden release of energy from a stellar mass compact object, such as a pulsar, magnetar, or accreting black hole (e.g.,~\citealt{Falcke&Rezzolla14,Lyubarsky14,Fuller&Ott15,Beloborodov17,Ioka&Zhang20,Sridhar+21b}).  The emission mechanism could be synchrotron maser emission at a relativistic magnetized shock \citep{Lyubarsky14,Beloborodov17,Metzger+19,Plotnikov&Sironi19}, fast magnetosonic waves generated by magnetic reconnection \citep{Philippov+19,Lyubarsky19,Lyubarsky20}, or curvature radiation resulting from charge starvation in a neutron star magnetosphere \citep{Kumar+17,Lu+20}, among other possibilities (e.g., \citealt{Wadiasingh2019,Kumar&Bosnjak20,Lyutikov21}).  However, one feature potentially common to a wide range of models is a power-law evolution of the burst properties relative to some zero point (corresponding to the onset of the energy release), which creates the observed emission as the peak of the spectral energy distribution of the sources sweeps down across the instrumental band-pass.  For example, in the synchrotron maser shock model (e.g., \citealt{Lyubarsky14,Beloborodov17,Margalit+20,Sironi+21})
the FRB emission drops in luminosity and frequency as the shock decelerates and probes lower densities at larger radii from the central engine \citep{Metzger+19,Margalit+19}.  Magnetospheric models for FRB emission could also predict power-law decay in the burst characteristics, for instance as a result of radius-to-frequency mapping as the emission region moves outwards in the magnetosphere (e.g., \citealt{Lyutikov2020}).

In this paper, we introduce a simple, physically-motivated `toy' model for the time-frequency structure of FRB emission, which we demonstrate is consistent with many of their observed features.  We then explore to what extent this burst structure could unify the dichotomy in FRB properties observed by CHIME \citep{Pleunis+21}.  As we will show, varying just one parameter of the model$-$the power-law index of the frequency drift rate$-$can transform a long and hence preferentially time-resolved burst into a (preferentially unresolved) short burst.  Insofar that the same drift-rate parameter determines the frequency bandpass over which the burst's signal-to-noise (SNR) is accumulated, this will act to preferentially narrow the observed time-integrated spectra of longer bursts in an SNR-limited sample, consistent with the CHIME dichotomy \citep{Pleunis+21}.  If confirmed by future work, this interpretation for the duration/spectral-width dichotomy would further refine the question of what distinguishes repeating from non-repeating FRB sources into one of how the toy model parameters map onto the diversity of the FRB central engines or their activity levels.   

This paper is organized as follow.  In Section \ref{sec:model} we introduce the toy model.  In Section \ref{sec:observables} we connect the model parameters to various FRB observables.  Section \ref{sec:implications} discusses implications of the model for burst detectability and for interpreting the CHIME dichotomy.  Section \ref{sec:conclusions} summarizes our conclusions.

\section{Toy Model of FRB Emission}
\label{sec:model}

\begin{figure}
    \centering
    \includegraphics[width=0.45\textwidth]{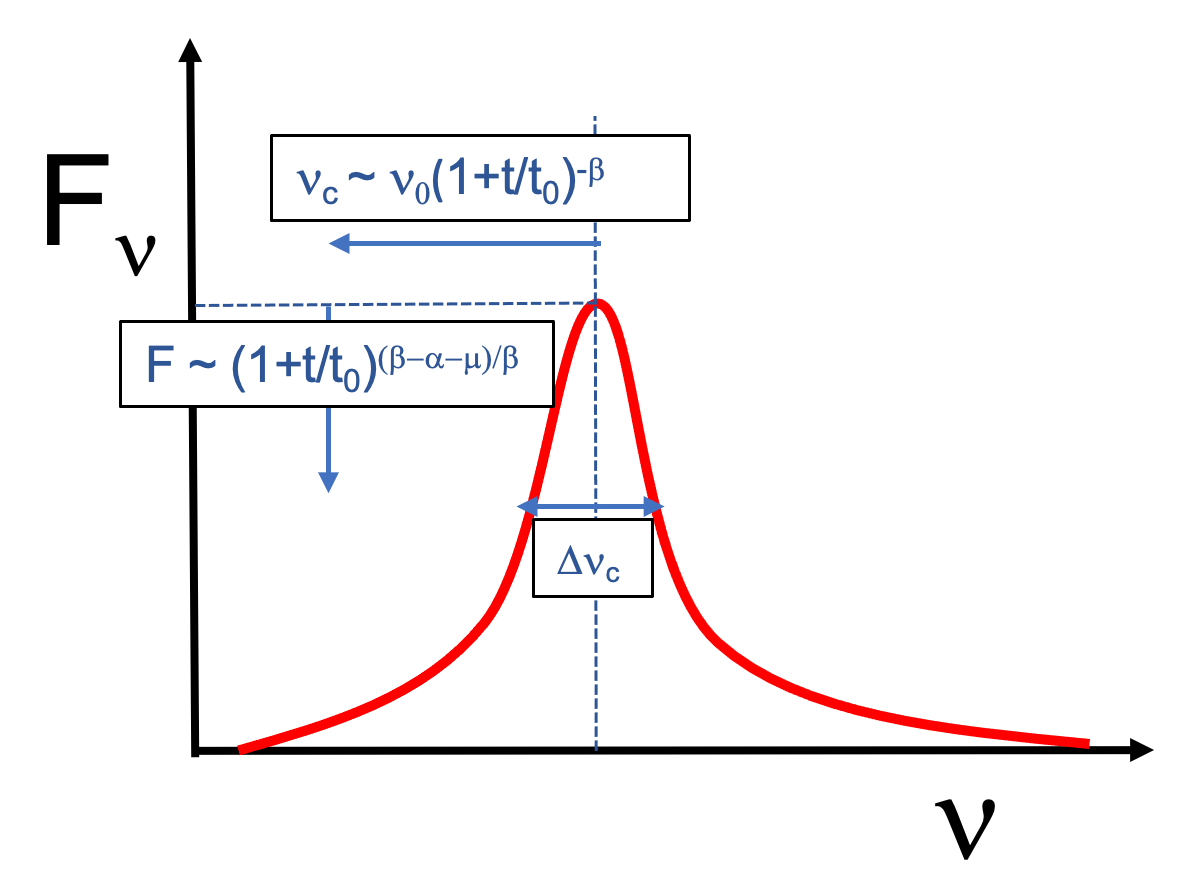}
    \caption{Schematic description of the model for the time-dependent spectral energy distribution of FRBs. }
    \label{fig:schematic}
\end{figure}

\begin{figure*}
    \centering
    \includegraphics[width=0.32\textwidth]{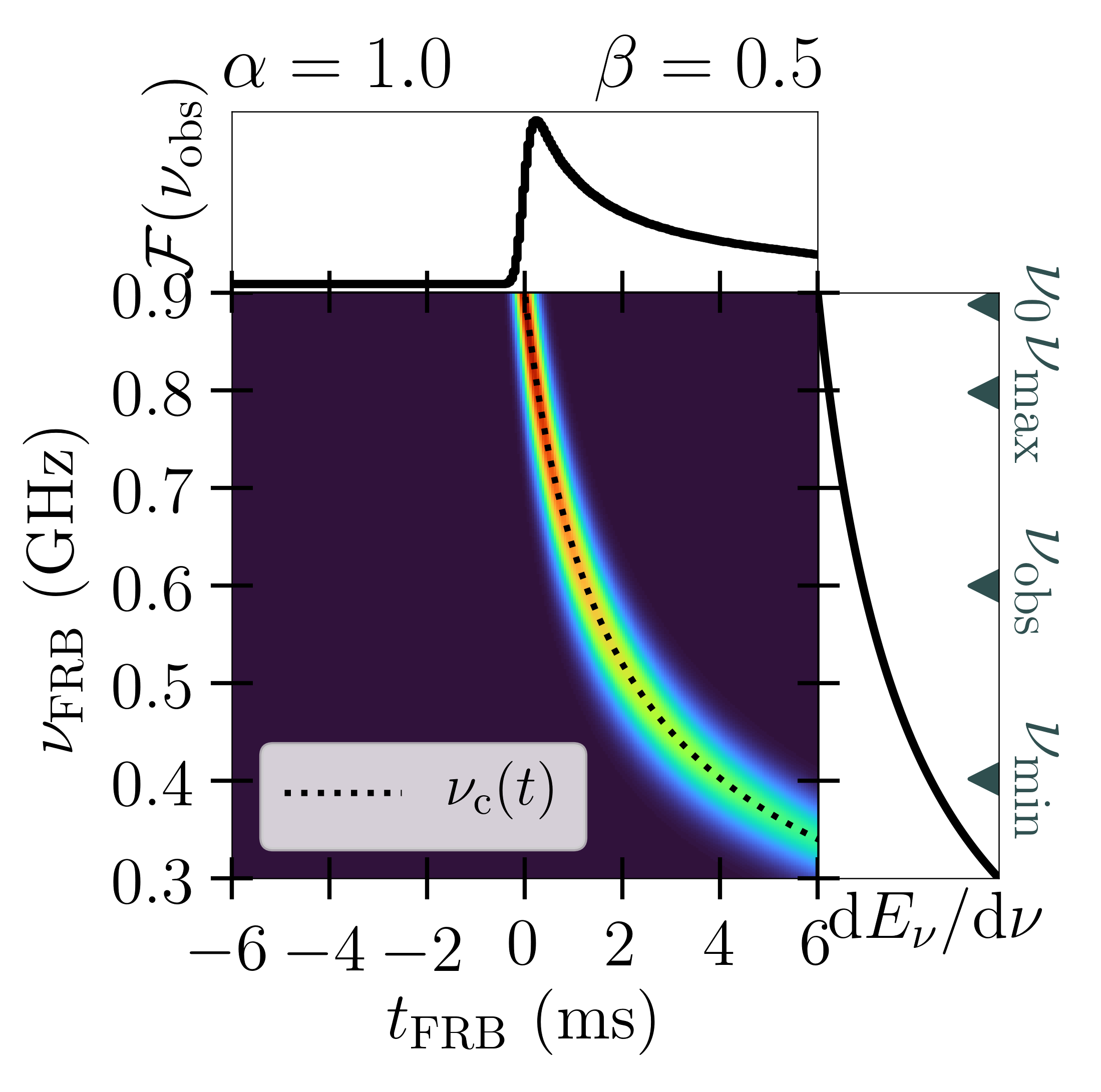}
    \includegraphics[width=0.32\textwidth]{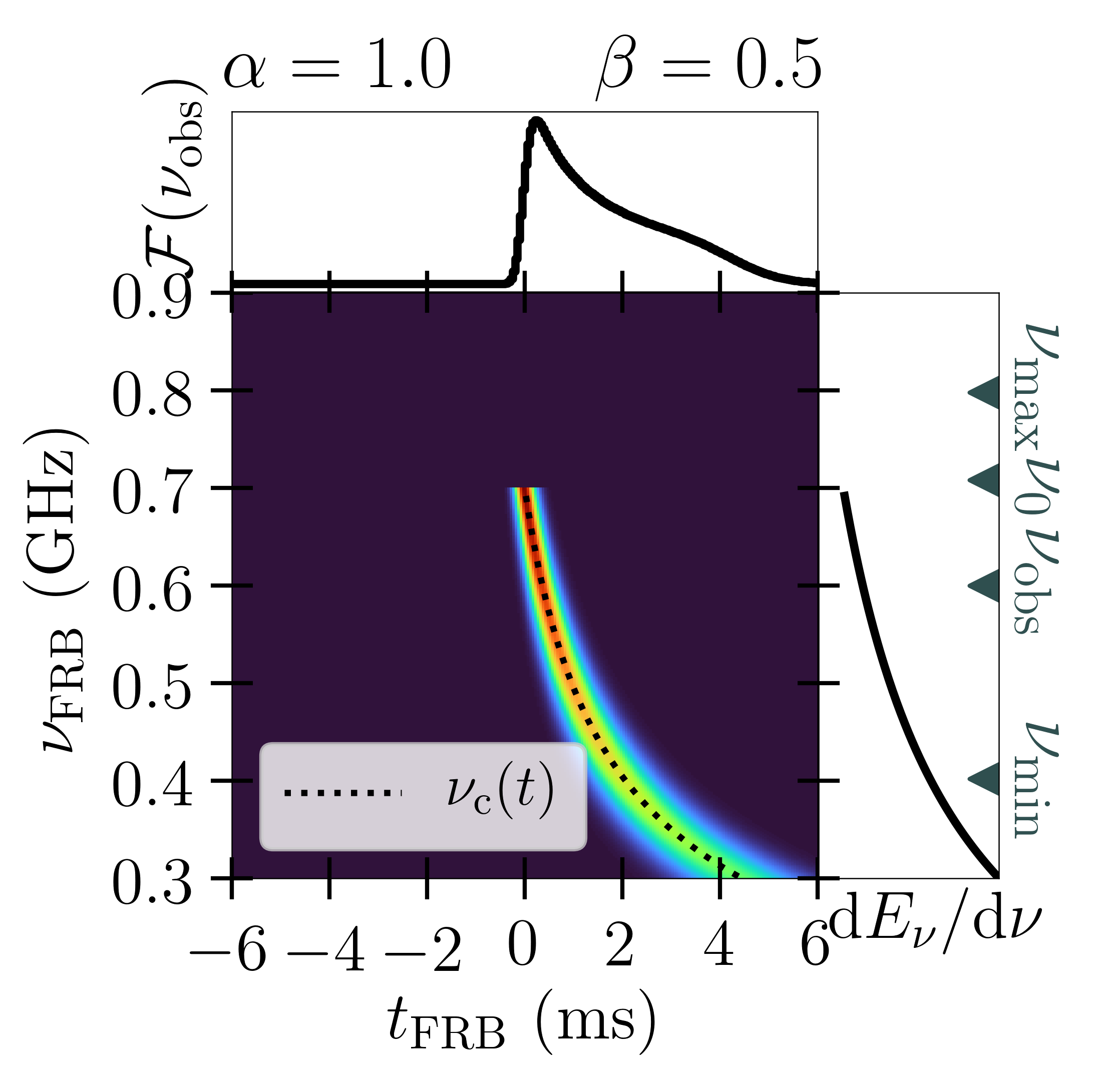}
    \includegraphics[width=0.32\textwidth]{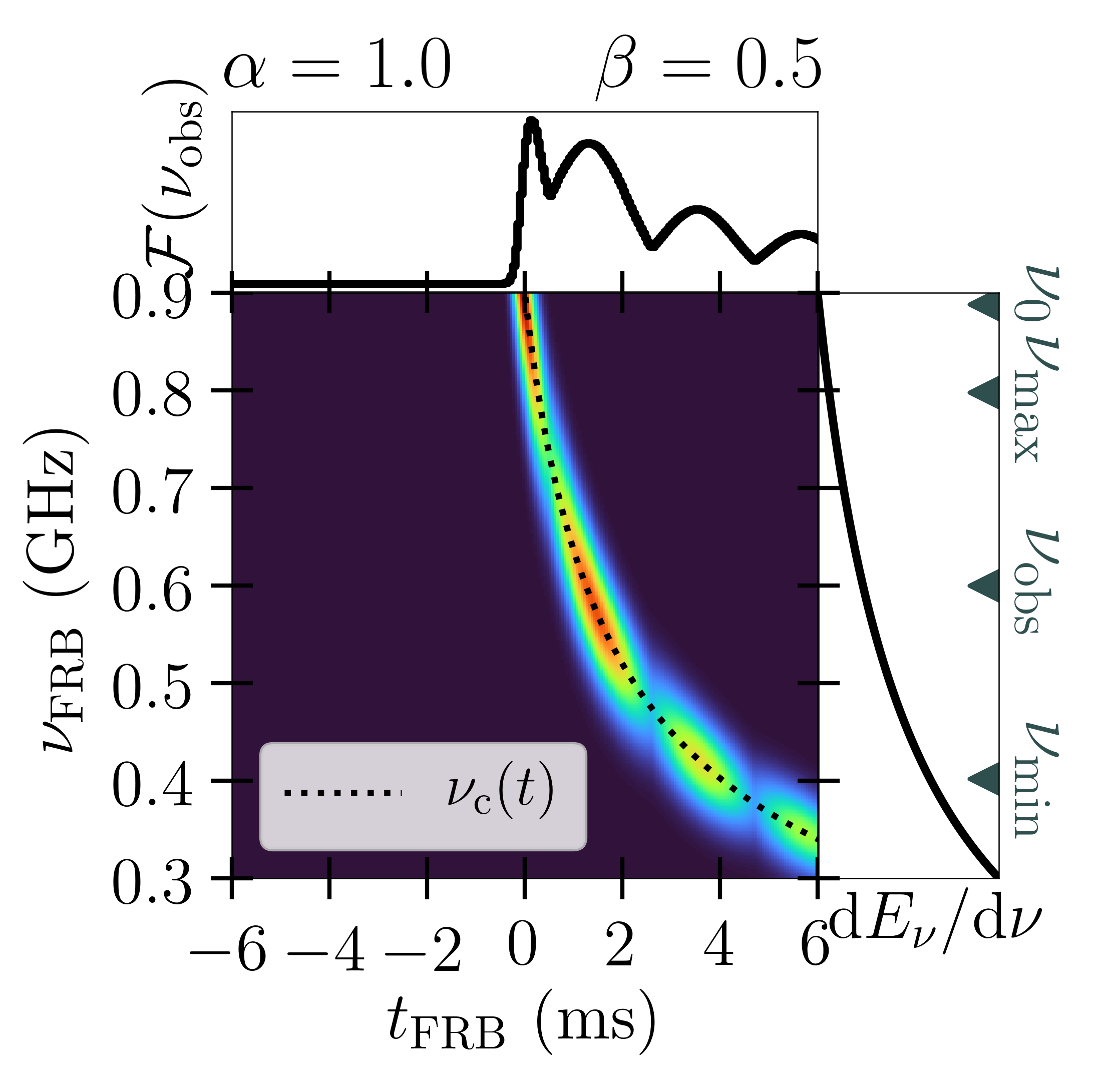}
    \caption{Synthetic time-frequency (``waterfall'') plots of $F_{\nu}(t,\nu)$ for an FRB characterized by $\alpha = 1$, $\beta = 0.5$, $\chi = 0.1$, $t_{\rm 0}=$ 1 ms.  The left panel shows an example in which the burst starts above the observing band ($\nu_0 > \nu_{\rm max}$), while the middle panel shows an example where the burst starts in-band ($\nu_{\rm min} < \nu_0 < \nu_{\rm max}$). The right panel shows an otherwise identical burst model to the left panel but for which we have applied a periodic Gaussian filter to the signal, mimicking possible propagation effects, which acts to break the signal into distinct sub-bursts.}
    \label{fig:waterfallschematic}
\end{figure*}

We introduce a parameterized model for the time-frequency structure of FRB emission (see Figure \ref{fig:schematic} for a schematic illustration).  Though originally motivated by shock scenarios, in which the free parameters are set by the properties of the ejecta from the central engine and of the external upstream medium, this model can in principle accommodate a wider range of emission mechanisms.  

\subsection{Single Component Burst Model}
\label{sec:simple}

The flux density of a simple, single component FRB evolves in frequency $\nu$ and time $t$ according to
\be
F_{\nu}(t) = F(t)\exp\left[-\frac{(\nu-\nu_{\rm c}(t))^{2}}{\Delta \nu_{\rm c}(t)^{2}}\right],
\label{eq:Fnu}
\ee
where $\nu_{\rm c}(t)$ and $\Delta \nu_{\rm c}(t)$ are the central frequency and width, respectively, and the normalization $F(t)$ is a function defined below.  Our particular choice of a Gaussian shape for the spectral ``envelope'' is somewhat arbitrary; however, the qualitative conclusions to follow will hold for more general spectral shapes insofar as they are narrowly peaked (i.e., $\Delta \nu_{\rm c} \ll \nu_{\rm c}$), consistent with time-resolved FRB observations (e.g., \citealt{Kumar+21}).  

Motivated by the observed downward drifting of frequency structure in FRB bursts (e.g., \citealt{Hessels+19}), the central frequency is assumed to decay as a power-law in time,
\be
\nu_{\rm c}(t) = \nu_0\left(\frac{t-\tilde{t}}{t_0}\right)^{-\beta},
\label{eq:nuc}
\ee
where $\tilde{t}$ is a singular time (usually corresponding to the point of energy release from the central engine) and $\nu_0$ the central frequency at some fiducial time $t_0$ after $\tilde{t}$.  Eqs.~(\ref{eq:Fnu}), (\ref{eq:nuc}) cannot be valid at all times, since in general this would lead to a divergence in the total burst energy.  To alleviate this, we assume the burst emission starts after a time $t = \tilde{t} + t_0$ at which point $\nu_{\rm c}(t_0) = \nu_0$.  For example, in shock scenarios, the start time $t_0$ could correspond to when the relativistic ejecta from the central engine has swept up a sufficient mass from the external medium to appreciably decelerate (Appendix \ref{app:shock}).  \citet{Beniamini&Kumar20} explore a similar setup, from which they explore how the measured burst duration at a given frequency is related to the total energy emitted in that band.   

Rewriting Eq.~(\ref{eq:nuc}) in terms of a new time variable, now measured starting at $\tilde{t} + t_0$, we find
\be
\nu_{\rm c}(t) = \nu_0\left(1 + \frac{t}{t_0}\right)^{-\beta}.
\label{eq:nuc2}
\ee
As discussed further below, there are two cases to consider (left and central panels of Fig.~\ref{fig:waterfallschematic}).  When the initial frequency $\nu_0$ is above the instrument band-pass, then a burst will contribute flux across the entire band-pass as $\nu_c$ sweeps downwards through it.  By contrast, when $\nu_0$ is within the instrument band-pass, the observed emission starts at $\nu_0$ and is restricted to frequencies $\nu < \nu_0$.

The frequency width of the emission envelope can in principle also evolve independently of the central frequency:
\be
\frac{\Delta \nu_{\rm c} }{\nu_{\rm c}} = \chi\left(1 + \frac{t}{t_0}\right)^{-\mu}.
\label{eq:width}
\ee
In most of the examples presented throughout this paper, we will take $\mu = 0$, and hence $\Delta \nu_{\rm c}/\nu_{\rm c} = \chi = constant$, where $\chi \ll 1$ to reproduce the narrow-band spectra of time-resolved bursts (e.g., \citealt{Farah+18,Michilli+18,Hessels+19, Day+20,Kumar+21}).  

The frequency-integrated total flux is also assumed to exhibit a power-law temporal decay, 
\be
\int_{0}^{\infty} F_{\nu}d\nu \approx \frac{\sqrt{\pi}}{2}\Delta \nu_{\rm c}(t) \cdot F(t) \equiv F_0 \left(1 + \frac{t}{t_0}\right)^{-\alpha},
\label{eq:integrated}
\ee
with a different index $\alpha,$ where $F_0$ is a constant.  From Eqs.~(\ref{eq:width}), (\ref{eq:integrated}) we deduce that
\begin{eqnarray}
F(t) &\approx& \frac{2}{\sqrt{\pi}}\frac{F_0}{\chi \nu_0}\left(1 + \frac{t}{t_0}\right)^{(\beta+\mu-\alpha)} \nonumber \\
&=& \frac{2}{\sqrt{\pi}}\frac{F_0}{\chi \nu_0}\left(\frac{\nu_{\rm c}(t)}{\nu_0}\right)^{(\alpha-\mu-\beta)/\beta}.
\label{eq:normalization}
\end{eqnarray}

Observed FRB spectra, as revealed by high time-resolution studies, are often extremely complex (e.g.~\citealt{Farah+18,Michilli+18,Day+20,Cho+20, Nimmo_20}) and do not conform in detail to the model description above, which is smooth in time and frequency.  However, it is possible that the observed fine-scale time-frequency structures are imprinted by propagation effects, such as plasma lensing (e.g., \citealt{Cordes+17,Simard&Ravi20,Sobacchi+21,BKN2021}), rather than being an intrinsic property of the emission.  In particular, the dissolution of the FRB light curve into a train of ``sub-bursts" (often roughly equally spaced in frequency)---suggestive of interference fringes---may well result from such effects (see also right panel of Fig.~\ref{fig:waterfallschematic}). Nonetheless, there are at least several bursts in which heretofore considered propagation effects were argued unlikely to be the result of the observed strong spectro-temporal-spectral variability \citep{Beniamini&Kumar20,LBK2021}. 

We do not attempt to model propagation effects in this paper. Rather, our goal is to describe the overall ``envelope'' of the burst's time structure, under the (admittedly strong) assumption that, in spite of any propagation-induced modulation, key features of the intrinsic emission are preserved and can be described within the framework of the toy model.  This assumption would be violated, for instance, if key features of the burst emission, such as the downwards frequency evolution, are also the result of propagation effects (e.g., additional dispersion-induced delay close to the source; \citealt{Tuntsov+21}).

\subsubsection{Model Parameters in Physical Emission Models}

Appendix \ref{app:shock} and Table \ref{tab:shock} summarizes the toy model parameters $\{\alpha, \beta \}$ predicted for synchrotron maser emission from a relativistic shock propagating into a stationary or sub-relativistically expanding external medium \citep{Metzger+19}.  For the typically considered case in which emission from the shock is observed on a timescale longer than that of the energy release from the central engine ($t \gtrsim \delta t$), the scenario predicts $\alpha \approx 1$ and $\beta \approx 0.2-0.7$, depending on the power-law index of the radial density profile of the upstream medium, $n(r) \propto r^{-k}$ ($k \approx 0-2$).  

However, a wider range of $\{\alpha$, $\beta\}$ values are permitted under other assumptions within the shock model, such as a relativistically expanding upstream medium \citep{Beloborodov19,Sridhar+21}.  For example, for internal shocks in the accelerating relativistic wind of a binary neutron star merger \citep{Sridhar+21}
\be  \alpha = \frac{4m-21}{2(m-5)} \in [1,1.75];\,\,\,\,\,\,\beta = \frac{6m-27}{4(m-5)} \in [1.9,3],
\label{eq:BNS}
\ee
where $\dot{M} \propto a^{-m}$ is the (theoretically uncertain) time-dependent mass-loss rate of the wind as a function of the shrinking binary semi-major axis $a$ and $m \in [5.5,7]$. 

Magnetospheric models, in which the frequency drift is set by the radius-to-frequency mapping typically predict $\beta \simeq 1$ (e.g., \citealt{Lyutikov2020}), as long as the magnetosphere is slowly rotating and for a wide range of dependencies of $\nu_{\rm c}$ on the dipole magnetic field. Faster (relativistically) rotating magnetospheres, result in a drift rate that can deviate significantly from a power-law dependence and which can also vary significantly between different bursts from the same engine.

\subsection{Multi-Component Burst Model}
\label{sec:complex}

\begin{figure*}
    \centering
    \includegraphics[width=0.45\textwidth]{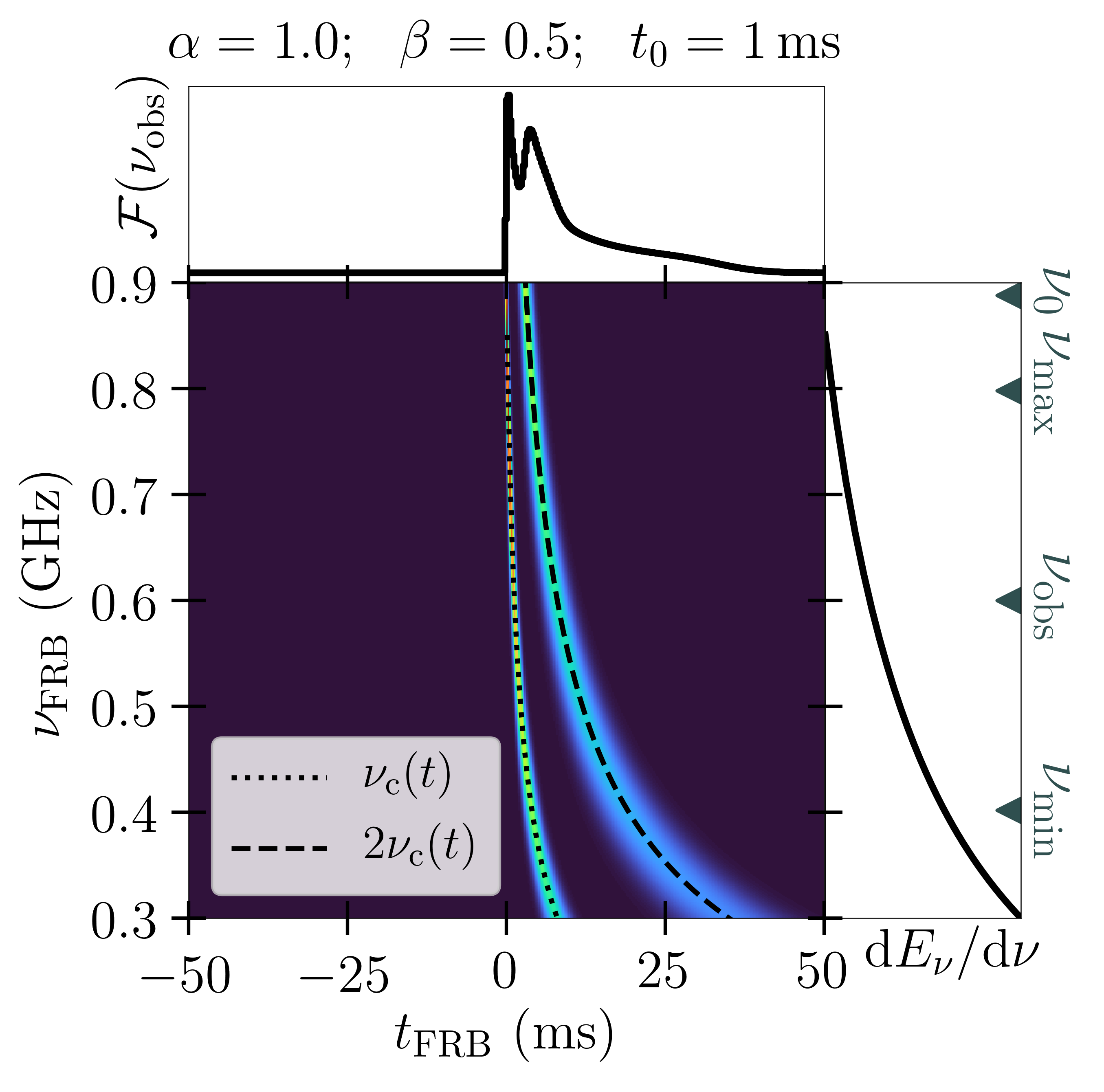}
    \hspace{1.5cm}
    \includegraphics[width=0.45\textwidth]{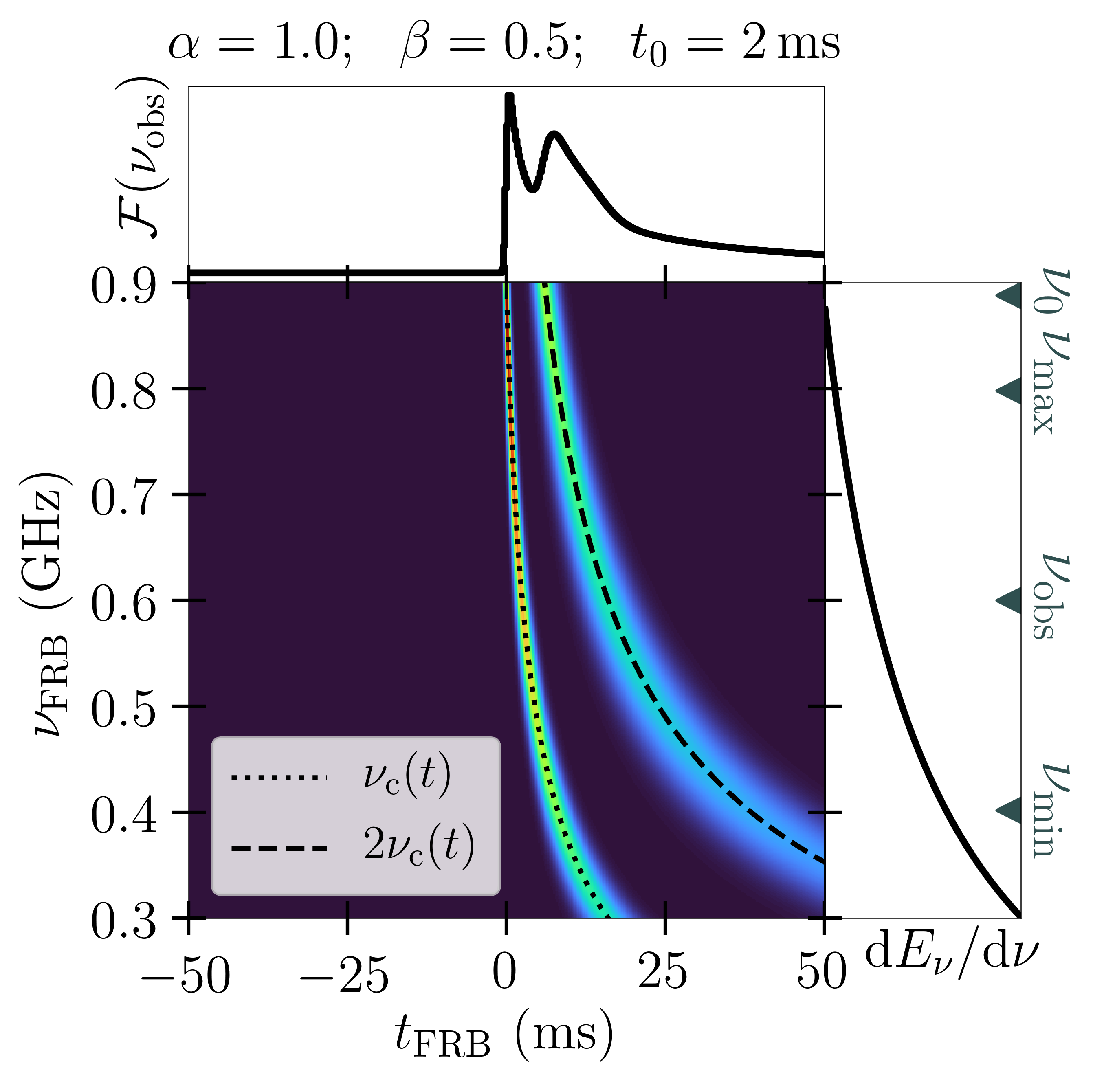}
    \caption{Synthetic time-frequency (``waterfall'') plots for an FRB with two sub-components with central frequencies $\nu_{\rm c}$ and $2\nu_{\rm c}$, according to the formalism in Section \ref{sec:complex}. The burst profiles are calculated for otherwise fixed values of $\alpha = 1$, $\chi = 0.1$, $\nu_0 = 0.9$\,GHz, $\beta = 0.1$ but varying values of $t_{\rm 0} = 1$\,ms (left), $t_{\rm 0} = 2$\,ms (right). This illustrates that the time separation between multiple sub-components increases with the duration of each component $\propto t_0$.}
    \label{fig:components}
\end{figure*}

Although our main focus in this paper is on the simple burst model presented in Section \ref{sec:simple}, possible generalizations to more complex light curves should be kept in mind.  In particular, FRB light curves composed of multiple separate light curve peaks can be accommodated within the same framework as above by summing multiple Gaussian components with distinct values of $F_{i}(t)$ and $\nu_{\rm c,i}$ (each obeying Eqs.~\ref{eq:nuc2}, \ref{eq:normalization} for the same values of $\alpha$ and $\beta$, but different initial frequencies $\nu_0 = \nu_{0,i}$), viz.~
\be
F_{\nu}(t) = \sum_{i}F_{i}(t)\exp\left[-\frac{(\nu-\nu_{\rm c,i}(t))^{2}}{\Delta \nu_{\rm c,i}(t)^{2}}\right].
\label{eq:Fnutot}
\ee
As each individual frequency component $\nu_{\rm c,i}$ sweeps through the observational band-pass, it will generate its own peak in the light curve.  Such a burst structure could be motivated, for instance, in shock models if harmonics of the synchrotron maser emission generate distinct peaks in the burst's spectral energy distribution (e.g., \citealt{Babul&Sironi20}), which appear as distinct burst components as each harmonic peak crosses the instrumental bandpass. In such cases, the separation between two sub-components increases as the burst width increases (see Fig.~\ref{fig:components} for an example).

\begin{figure*}
    \centering
    \includegraphics[width=0.45\textwidth]{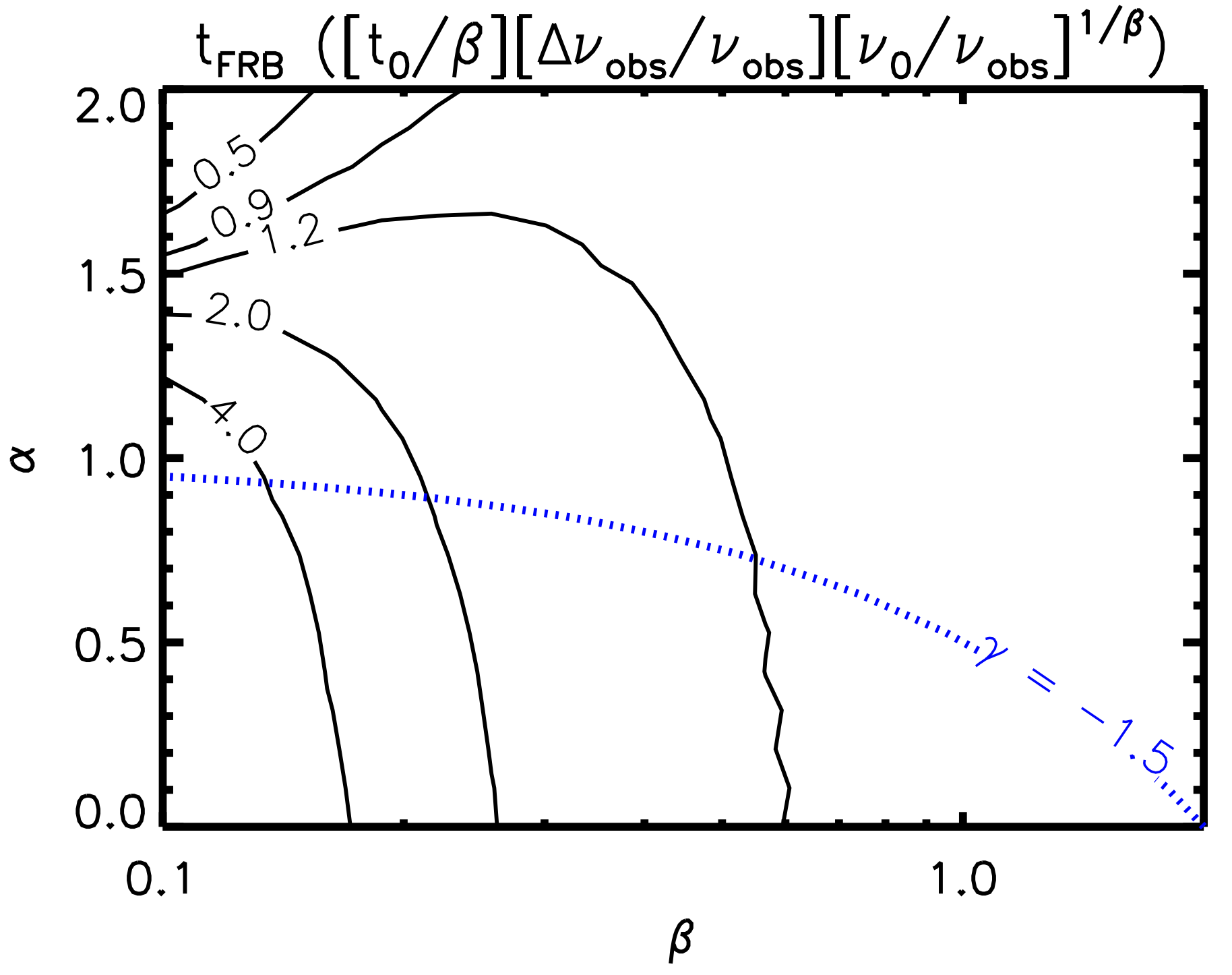}\hspace{0.9cm}
    \includegraphics[width=0.45\textwidth]{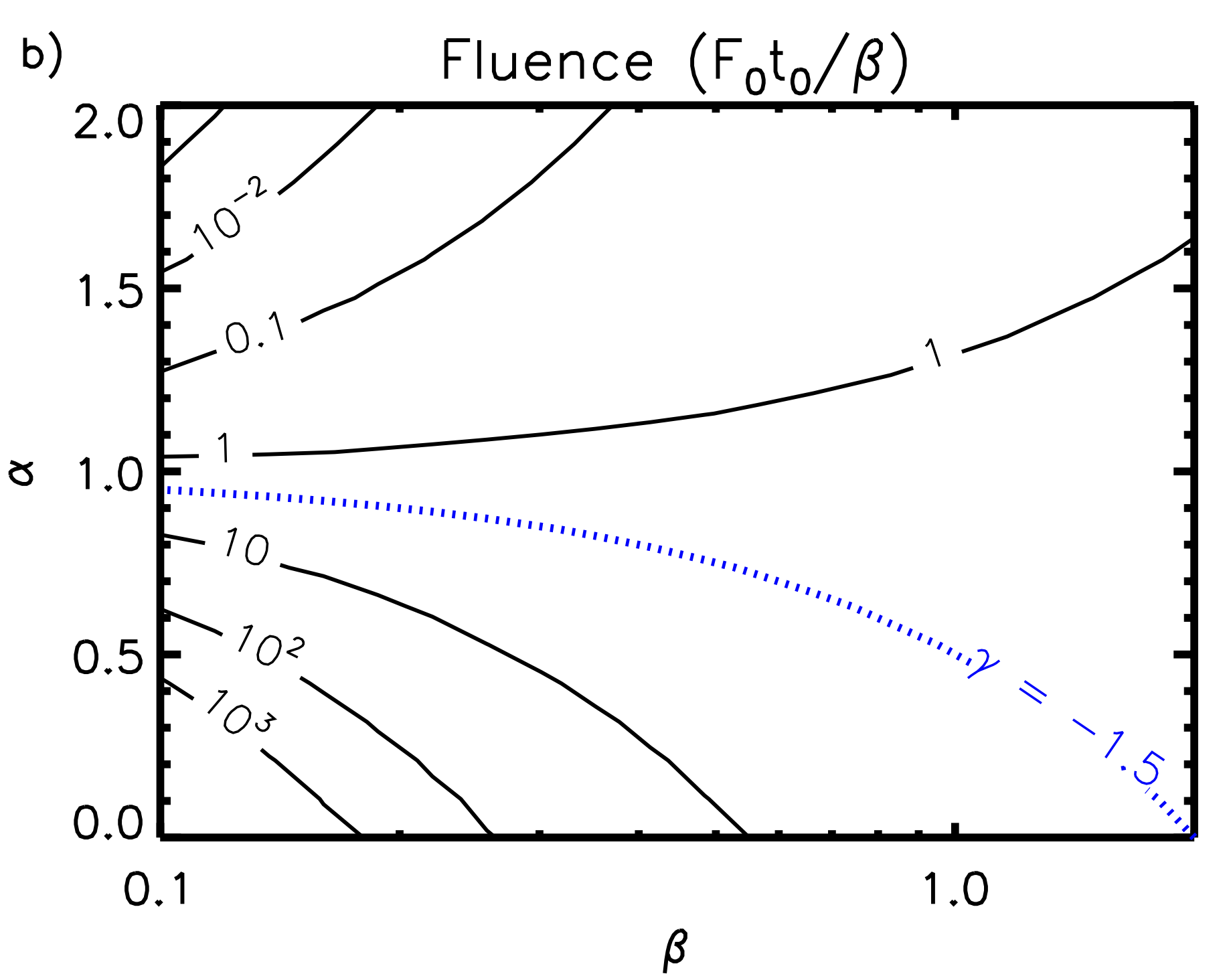}
    \includegraphics[width=0.45\textwidth]{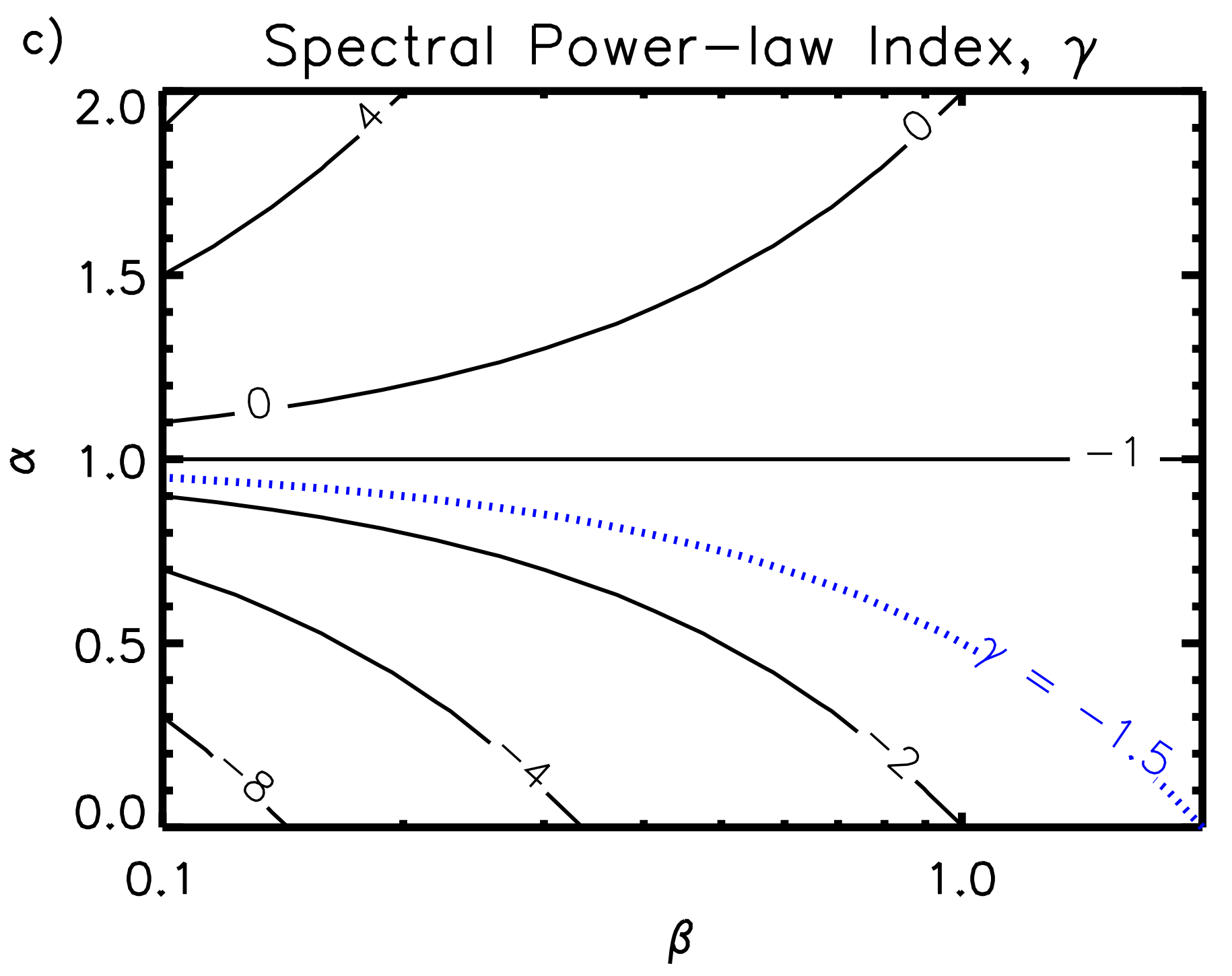}\hspace{0.9cm}
    \includegraphics[width=0.45\textwidth]{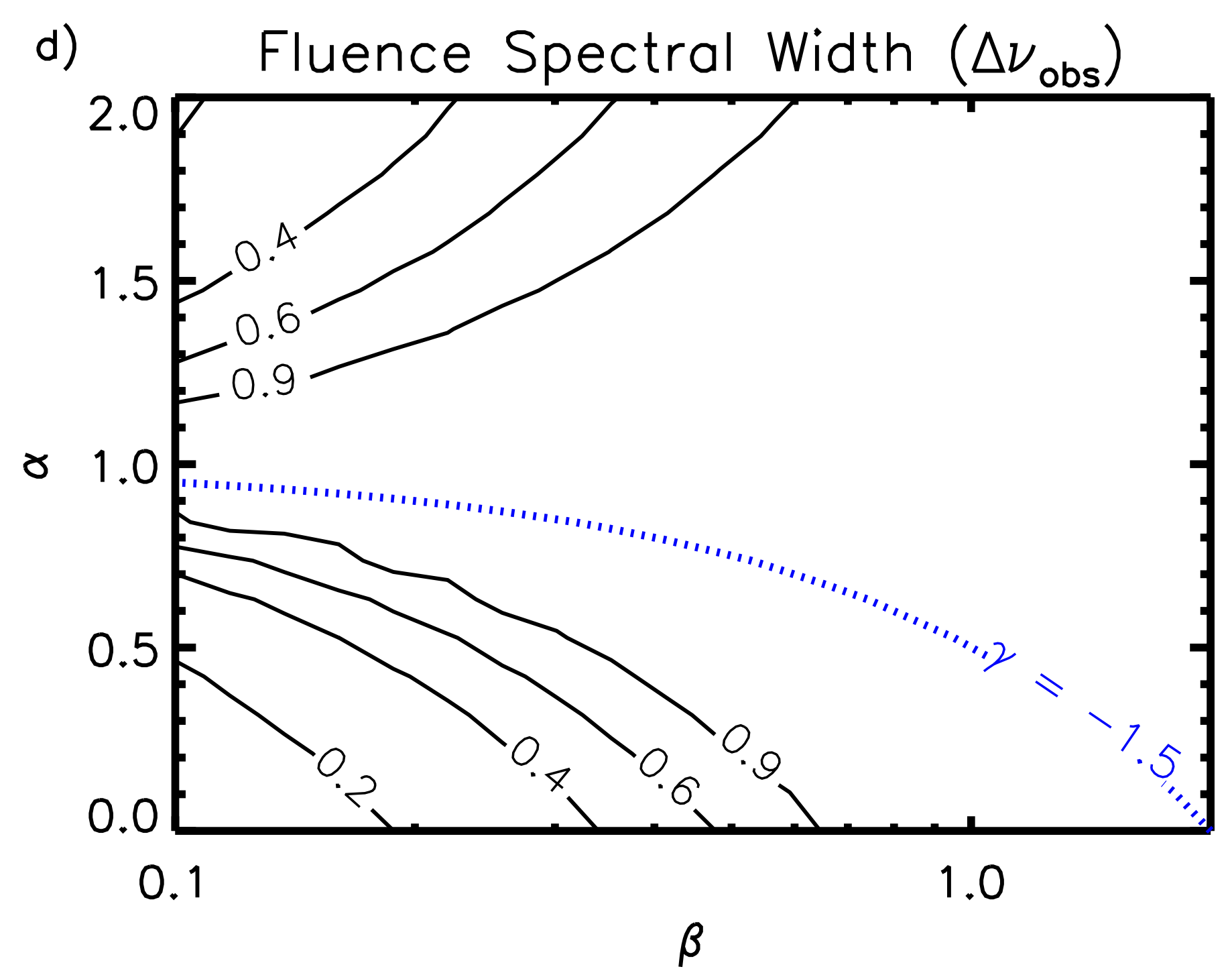}
    \includegraphics[width=0.45\textwidth]{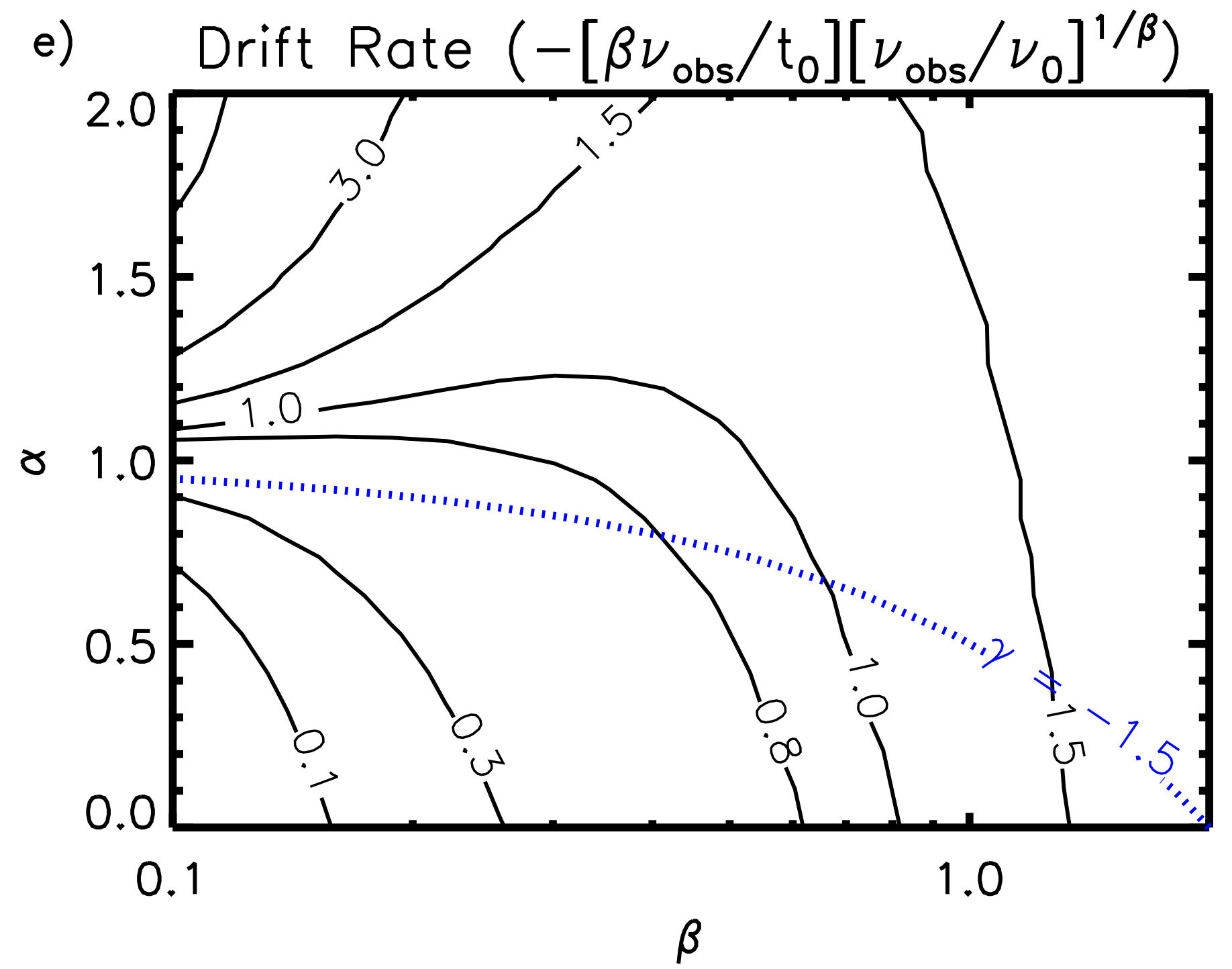}\hspace{0.9cm}
    \includegraphics[width=0.45\textwidth]{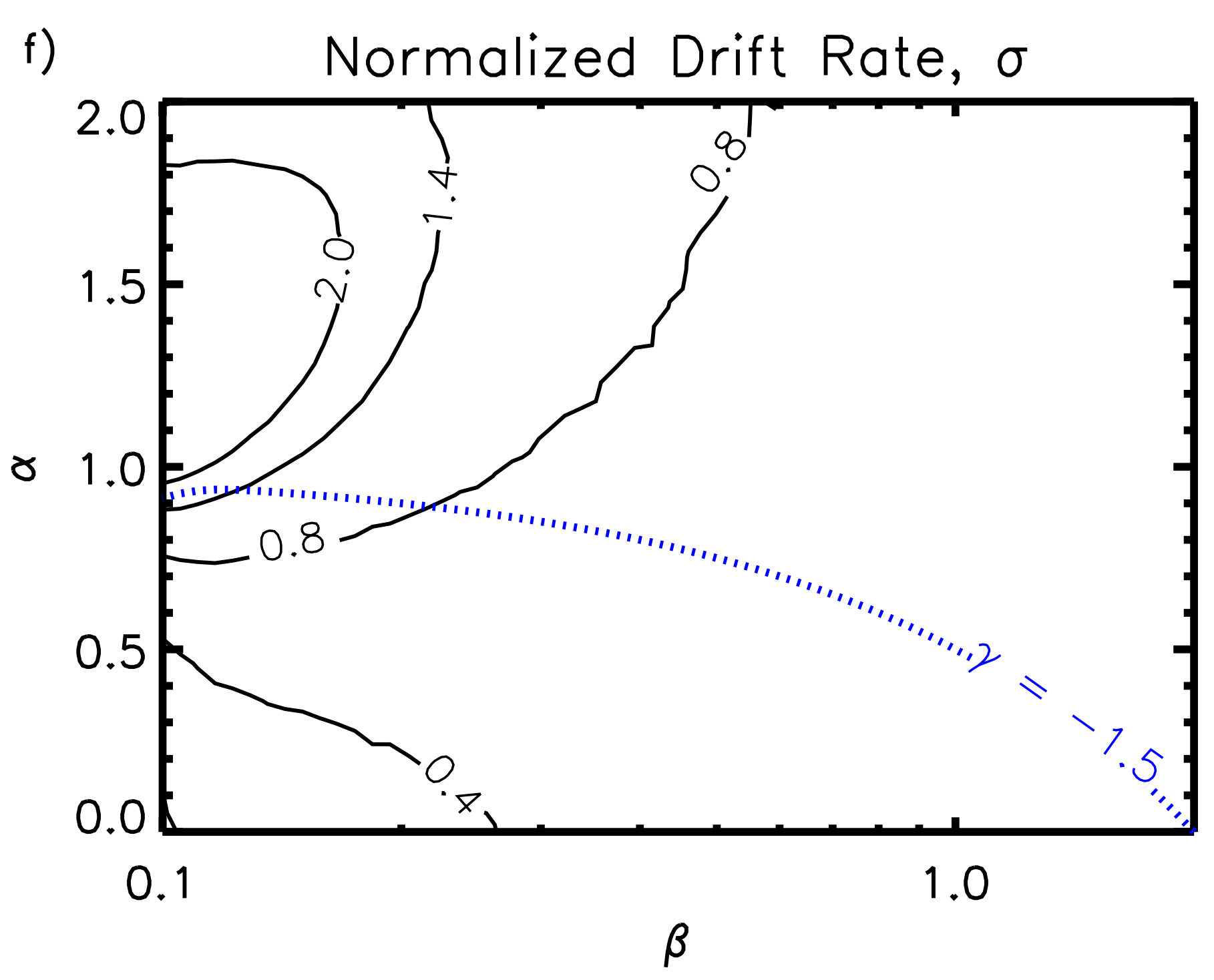}
    \caption{Observable properties of the FRB as a function of the power-law indices $\alpha$ and $\beta$, calculated directly from Eq.~(\ref{eq:Fnu}) and normalized to the analytic estimates presented in the text (which typically work best around $\alpha \approx 1$).  We have assumed: $\mu = 0$, $\chi = 0.1$, $\nu_0 = 2\nu_{\rm max}$; and an instrument with temporal resolution $\Delta t_{\rm res} \ll t_0$ and band-pass $\Delta \nu_{\rm obs}/\nu_{\rm obs} \approx 0.67$ matching that of CHIME.  Quantities shown include: a) burst duration $t_{\rm FRB}$, defined as the time interval over which 90\% of the burst fluence is accumulated (normalized to Eq.~\ref{eq:tobs}); b) burst fluence (normalized to Eq.~\ref{eq:fluence}); c) time-integrated spectral power-law index (agrees well with Eq.~\ref{eq:gamma}); d) spectral width over which 50\% of the burst fluence is received, normalized to the instrument band-pass; d) frequency drift rate, measured at the point of 50\% accumulated fluence (normalized to Eq.~\ref{eq:nucdot}); e) dimensionless frequency drift rate (Eq.~\ref{eq:betaben}).  The dotted blue contour in each figure shows the average spectral index $\gamma \simeq -1.5$ found for a sample of ASKAP bursts by \citet{Macquart+19}.}
    \label{fig:observables}
\end{figure*}

\section{Burst Observables}
\label{sec:observables}

We now describe how the hypothesized time-frequency structure (Eq.~\ref{eq:Fnu}) of the burst maps onto observed FRB properties.  In what follows, we consider the emission as detected by an idealized telescope with sensitivity in the frequency range $\nu \in [\nu_{\rm min},\nu_{\rm max}$], with a central frequency $\nu_{\rm obs} = (\nu_{\rm min} + \nu_{\rm max})/2$ and total bandwidth $\Delta \nu_{\rm obs} = \nu_{\rm max}-\nu_{\rm min} \lesssim \nu_{\rm obs}$.  When necessary to specify, the instrumental time resolution is taken to be $\Delta t_{\rm res}$ ($\simeq 1$ ms for CHIME).  

Figure \ref{fig:observables} shows contour plots depicting the result of a direct numerical evaluation of the various burst properties described below, as a function of the model power-law indices $\alpha$ and $\beta$.  Many of the results in Fig.~\ref{fig:observables} have been normalized to analytic estimates derived below.  In this example we assume the emission begins above the top of the instrumental band-pass ($\nu_0 = 2\nu_{\rm max}$; similar to the example shown in the left panel of Fig.~\ref{fig:waterfallschematic}), though qualitatively similar results would follow if $\nu_0$ were to instead occur within the instrumental band-pass (as expected particularly for longer bursts that concentrate their SNR in a narrow frequency range about $\nu_0-$see Section \ref{sec:implications}). 
\subsection{Frequency Drift Rate}
\label{sec:drift}

Evolution of the burst frequency structure occurs as $\nu_{\rm c}$ crosses down through the observing band.  The drift rate as measured around frequency $\nu_{\rm c}$ is given by
\be
\dot{\nu_{\rm c}}   = -\frac{\beta \nu_{\rm c}}{t_0}\left(\frac{\nu_{\rm c}}{\nu_0}\right)^{1/\beta} \underset{\nu_{\rm c} \approx \nu_{\rm obs}}= -\frac{\beta \nu_{0}}{t_0}\left(\frac{\nu_{\rm obs}}{\nu_0}\right)^{(\beta+1)/\beta},
\label{eq:nucdot}
\ee
where in the second line we have taken $\nu_{\rm c} \approx \nu_{\rm obs}$ at the center of the band-pass.

For values of $\beta > 0$, the drift rate is negative (``sad trombone'') and its magnitude increases with radio frequency $\nu_{\rm obs}$, consistent with observations of the well-studied repeating source FRB 121102 \citep{Hessels+19,Josephy+19,CHIME+20,Caleb_20}.  Fitting the drift rates of distinct bursts from FRB 121102 measured across a range of radio frequencies from 0.6 GHz to 6.5 GHz, \citet{Caleb_20} find a linear dependence $\dot{\nu} \propto \nu_{\rm obs}$ (see also \citealt{Josephy+19}), which from Eq.~(\ref{eq:nucdot}) would require $\beta \gg 1$.  However, large scatter is observed in the drift rate at a fixed observing frequency (e.g., 1.4 GHz) for different bursts from a single repeating source.  Given that drift rates have not, to our knowledge, been measured simultaneously for the same burst at different frequencies, the value of $\beta$ cannot yet be precisely constrained.

As shown in Figure \ref{fig:observables}e, for $\alpha \sim 1$, Eq.~(\ref{eq:nucdot}) does a good job of reproducing the measured drift rate, as measured at the epoch of half accumulated fluence.  However, for larger(smaller) values of $\alpha$, the burst fluence peaks at the (top)bottom of the instrumental band-pass (Section \ref{sec:SED}) and hence the drift rate $\dot{\nu}_{\rm c} \propto \nu^{\frac{\beta+1}{\beta}}$ defined this way becomes very sensitive to $\beta$ for $\beta \ll 1$.  

\subsection{Burst Duration} 
\label{sec:duration}
One commonly used definition of the duration of a burst is the timescale over which the majority of its fluence is received.  This is roughly given by the time required for the (e.g., Gaussian) emission envelope to sweep down across the instrument band-pass $\Delta \nu_{\rm obs}$, 
\be
t_{\rm FRB} \sim \frac{\Delta \nu_{\rm obs}}{|\dot{\nu_{\rm c}}|} \underset{\nu_{\rm c} \approx \nu_{\rm obs}}\approx \frac{t_{\rm 0}}{\beta}\left(\frac{\Delta \nu_{\rm obs}}{\nu_{\rm obs}}\right)\left(\frac{\nu_{\rm obs}}{\nu_0}\right)^{-1/\beta},\,\text{single\,burst},
\label{eq:tobs}
\ee
where we have used Eq.~(\ref{eq:nucdot}).  As we shall discuss in Section \ref{sec:implications}, the burst duration is shorter than Eq.~(\ref{eq:tobs}) (typically by a factor $\sim \beta$ when $\beta \ll 1$) if it is defined as the time to accumulate most of the SNR instead of the fluence.  

On the other hand, for more complex bursts generated by multiple downward-drifting components in the SED (Eq.~\ref{eq:Fnutot}) spaced over a total frequency range $\sim \mathcal{N}\nu_{\rm c}$, where $\mathcal{N} \gtrsim 1$\footnote{In shock scenarios, for example, values of $\mathcal{N} \sim few$ are expected if each burst peak corresponds to an individual harmonic of the synchrotron maser emission \citep{Babul&Sironi20}.}, the total burst duration (as defined from the first to last peak of comparable fluence) can be considerably longer:
\begin{eqnarray}
&& t_{\rm FRB} \sim \mathcal{N} \cdot \frac{\Delta \nu_{\rm obs}}{\dot{\nu_{\rm c}}} \nonumber \\
& \underset{\nu_{\rm c}\approx \nu_{\rm obs}}\approx& \frac{\mathcal{N}t_0}{\beta}\left(\frac{\Delta \nu_{\rm obs}}{\nu_{\rm obs}}\right)\left(\frac{\nu_{\rm obs}}{\nu_0}\right)^{-1/\beta},\,\text{multi-peaked burst}.\nonumber \\
\label{eq:tobs2}
\end{eqnarray}
Note that in a multi-component burst, the total burst duration scales with the model parameters in the same way as the temporal separation between the individual light curve peaks (Fig.~\ref{fig:components}).  This behavior is consistent with observations by CHIME \citep{Pleunis+21}, provided the sub-bursts arise from distinct spectral components drifting down across the instrumental bandpass (instead of a single-component burst ``broken apart'' by propagation effects).

As shown in Figure \ref{fig:observables}a, Eq.~(\ref{eq:tobs}) does a reasonable job of reproducing the duration over which 90\% of the burst fluence is accumulated.  To order of magnitude, we have $t_{\rm FRB} \sim t_0/\beta$ for bursts that begin within or just above the instrumental band-pass (i.e., $\nu_0 \sim \nu_{\rm obs}$).  However, for bursts that instead start far above the observing frequency (i.e., $\nu_0 \gg \nu_{\rm obs}$) the burst duration $t_{\rm FRB}$ can be $\gg t_0$ for $\beta \ll 1$.  Also note that the duration of a given burst $\propto \nu_{\rm obs}^{-1/\beta}$ is significantly shorter if the burst is observed at higher radio frequencies (even once accounting for the effects of scattering during propagation to Earth), consistent with observations \citep{Michilli+18,Li+21}.

The burst duration and frequency drift rate can be combined into a dimensionless quantity:
\be
\sigma \equiv \frac{|\dot{\nu}_{\rm c}|t_{\rm FRB}}{\nu_{\rm obs}}
\label{eq:betaben}
\ee
Based on a sample of repeating FRBs with measured frequency drifts, \citet{Margalit+19} found values $\sigma \approx 0.1-0.4$ (what they define as ``$\beta$'').  Likewise, \citet{Chamma+21} find $\sigma \sim 0.1$ for three separate repeater sources (see their Fig.~1).  As shown in Fig.~\ref{fig:observables}f, this would appear to favor a parameter space around $\alpha \approx 0$ and $\beta \lesssim 0.2$ if the total burst duration is that of a single burst.  However, in cases when $t_{\rm FRB}$ represents the total duration of a train of closely spaced burst components (Eq.~\ref{eq:tobs2}), then the predicted value of $\sigma$ is correspondingly larger than shown in Fig.~\ref{fig:observables} by a factor $\mathcal{N} \gtrsim few$ and the permitted parameter space of $\{\alpha,\beta\}$ expands considerably.  Furthermore, the true burst duration entering Eq.~(\ref{eq:betaben}) will be shorter than estimated in Fig.~\ref{fig:observables} (and hence $\sigma$ over-estimated) if it is defined as the time over which the majority of the burst's SNR is accumulated instead of its fluence (Section \ref{fig:SNR}).

\subsection{Time-Integrated Spectral Energy Distribution} 
\label{sec:SED}

If the instrumental time resolution is sufficiently poor, $\Delta t_{\rm res} \gg t_{\rm FRB}$, it is not possible to measure the spectral peak sweeping down across the observing band-pass because the latter occurs on a time $t_{\rm FRB} \sim |\Delta \nu_{\rm obs}/\dot{\nu_{\rm c}}|$ shorter than $\Delta t_{\rm res}$.  In such cases, only the {\it time-integrated} flux, or spectral energy distribution (SED), can be measured.  For a single burst, this can be estimated as
\begin{eqnarray}
\frac{dE_{\nu}}{d\nu} &\approx& \int_{0}^{\infty} F_{\nu \approx \nu_{\rm c}} dt \nonumber \\
 &\approx&  F(\nu_{\rm c})\left|\frac{\Delta \nu_{\rm c}}{\dot{\nu}_{\rm c}}\right| \underset{\nu_{\rm c} \approx \nu}\approx  \frac{2}{\sqrt{\pi}}\frac{F_{0}t_0}{\beta \nu_0}\left(\frac{\nu}{\nu_{0}}\right)^{\frac{\alpha-\beta-1}{\beta}}, 
\label{eq:SED}
\end{eqnarray}
where we have used Eq.~(\ref{eq:normalization}).  For temporally-unresolved bursts the instrument thus measures a {\it power-law} spectrum,
\be \frac{dE_{\nu}}{d\nu} \propto \nu^{\gamma};\,\,\,\,\,
\gamma \equiv \frac{(\alpha-\beta-1)}{\beta}.
\label{eq:gamma}
\ee 
This power-law SED is confirmed by direct integration of Eq.~(\ref{eq:Fnu}) in Figure \ref{fig:observables}c, which shows that $\gamma \approx -1$ obtains across the parameter space straddling $\alpha \approx 1$.  This predicts $\nu (dE_{\nu}/d\nu) \sim const$, i.e.~approximately equal fluence per logarithmic frequency interval, which improves the likelihood of multi-band FRB detections (e.g., \citealt{Majid+20,Pearlman+20}).  Based on a sample of 23 ASKAP bursts, \citet{Macquart+19} found a mean power-law index $\gamma = -1.5^{+0.2}_{-0.3}$ from 23 ASKAP FRBs (shown as a blue dotted line in Fig.~\ref{fig:observables}), which according to Eq.~(\ref{eq:gamma}) would imply $\alpha \simeq 1-0.5\beta \lesssim 1$.

On the other hand, for $\alpha >1$($\alpha < 1$) the value of $\gamma$ is much greater than (much less than) $-1$, particularly for low values of $\beta \ll 1$.  This same parameter range acts to concentrate the burst fluence over a narrow frequency range $\Delta \nu_E < \Delta \nu_{\rm obs}$, near the top or bottom of the instrumental band-pass (Fig.~\ref{fig:observables}d; see below), respectively, or around $\nu_0$ when the burst starts within the instrumental band-pass ($\nu_0 < \nu_{\rm max}$). 

For a complex burst composed of multiple downward drifting frequency components (Eq.~\ref{eq:Fnutot}), the time-integrated SED is just the sum of the contributions from each component.  If each burst component shares the same parameter values (i.e., $\alpha$, $\beta$), then the time-integrated spectrum will remain a power-law with the same index $\gamma$ (Eq.~\ref{eq:gamma}).  

\subsection{Burst Fluence}  
\label{sec:fluence}

The total fluence of the burst is obtained by integrating Eq.~(\ref{eq:SED}) across the relevant frequency range
\begin{eqnarray}
E &=& \int_{\nu_{\rm min}}^{\nu_{\rm max}'} \frac{dE_{\nu}}{d\nu}d\nu  \nonumber \\
&\underset{\alpha \ne 1}\approx& \frac{2}{\sqrt{\pi}}\frac{F_0 t_0}{(\alpha-1)}\left[\left(\frac{\nu_{\rm max}'}{\nu_0}\right)^{(\alpha-1)/\beta} - \left(\frac{\nu_{\rm min}}{\nu_0}\right)^{(\alpha-1)/\beta}\right] \nonumber \\
&& \underset{\alpha > 1}\approx \frac{2}{\sqrt{\pi}}\frac{F_0 t_0}{(\alpha-1)}\left(\frac{\nu_{\rm max}'}{\nu_0}\right)^{(\alpha-1)/\beta}   \label{eq:fluence}
\end{eqnarray}
where the maximum observed frequency,
\be \nu_{\rm max}' = {\rm min}[\nu_{\rm max},\nu_0],
\ee
is the top of the band-pass $\nu_{\rm max}' = \nu_{\rm max}$ if the burst starts above the band-pass (i.e., $\nu_0 > \nu_{\rm max}$) or $\nu_{\rm max}' = \nu_0$ if the burst starts within the band-pass (i.e., $\nu_{\rm min} < \nu_0 < \nu_{\rm max}$).  As shown in Figure \ref{fig:observables}b, the naive estimate $E \sim F_0 t_{\rm FRB} \sim F_0 t_0/\beta$ is accurate only around $\alpha \sim 1$.  Fig.~\ref{fig:observables}d shows the spectral width, $\Delta \nu_E$, over which half the burst fluence is accumulated, normalized to the instrumental band-pass $\Delta \nu_{\rm obs}$

\subsection{Summary of Burst Properties}

\begin{table*}
\centering
\caption{Summary of Burst Observables (for $\mu = 0$ and $\nu_0 \gtrsim \nu_{\rm obs}$)}
\begin{tabular}{ccc}
\hline
Burst Property & Controlling Parameters & Eq. \\
\hline
Onset Time & $\tilde{t}$ & \ref{eq:nuc} \\
Burst Duration, $t_{\rm FRB}$ & $t_0, \beta, \nu_0$ & \ref{eq:tobs} \\
Frequency Drift Rate, $\dot{\nu}_{\rm c}$ & $\beta, t_0, \nu_0$ & \ref{eq:nucdot}\\
Total Fluence, $E$ & $F_0, t_0, \alpha, \beta, \nu_0$ & \ref{eq:fluence} \\
Power-law Index of Time-Integrated SED, $\gamma$ & $\alpha, \beta$ & \ref{eq:gamma}\\
Time-Resolved Instantaneous Flux Bandwidth, $\Delta \nu_{\rm c}$  & $\chi, \nu_0$ & \ref{eq:width} \\
Time-Integrated Maximum SNR Bandwidth, $\Delta \nu_{\rm SNR}$ & $F_0, t_0$, $\alpha$, $\beta$, $\nu_0$ & \ref{eq:SNR1}, \ref{eq:SNR2}
\end{tabular}
\label{tab:summary}
\begin{flushleft}
\end{flushleft}
\end{table*}

In summary, for $\mu = 0$ and $\chi \ll 1$ (intrinsically narrow spectra), the FRB model is described by six free parameters: $\{\tilde{t}, t_0, \nu_0, F_0, \alpha, \beta \}$.  The singular time $\tilde{t}$ is related trivially to the epoch marking the beginning of the burst and hence is not normally considered a parameter of intrinsic interest (except in relation to other bursts).  For the typical case in which $\nu_0$ is within or just above the observing band-pass, the parameters $\beta$ and $t_0$ control the burst drift rate and the burst duration.  Along with $\beta$, the value of $\alpha$ determines the power-law slope of the time-integrated SED, $\gamma$ (Eq.~\ref{eq:gamma}).  The burst fluence $E$ mainly depends on the product of $F_0 t_0$, though the values of $\alpha$ and $\beta$ also enter for $\alpha \ne 1$.  A burst with more complex time structure can be built up by adding together many individual bursts (corresponding to individual narrow features in the intrinsic SED), with in general different values for $\{\nu_0, F_0\}$ but potentially the same values for $\{\tilde{t}, t_0, \alpha, \beta\}$.  

\section{Burst Detectability and the CHIME Dichotomy}
\label{sec:implications}

\begin{figure}
    \includegraphics[width=0.45\textwidth]{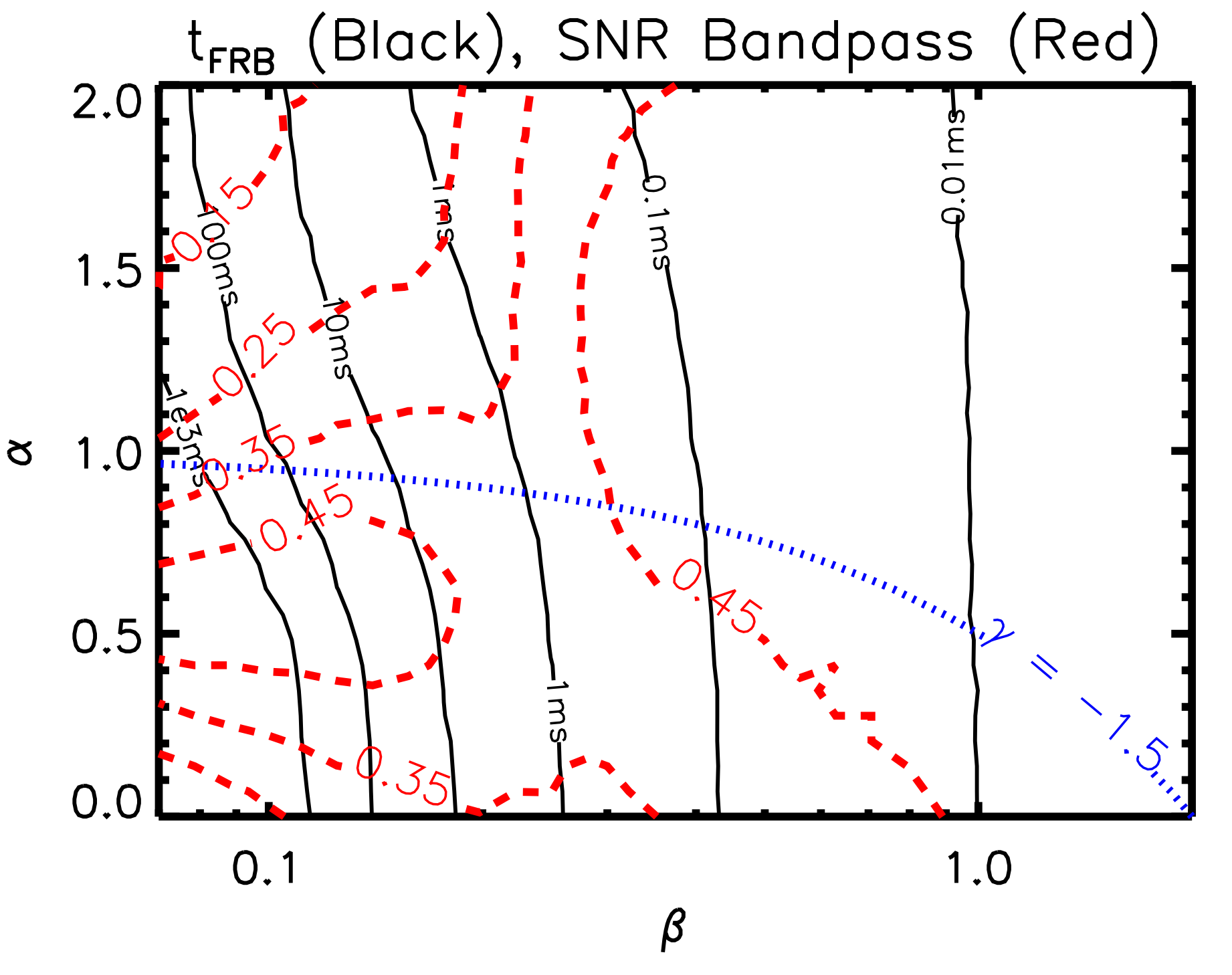}
    \caption{Black solid lines show the burst duration in milliseconds as a function of the power-law indices $\alpha$ and $\beta$, while dashed red contours show the fractional bandwidth over which half of the burst SNR is accumulated.  Here, we define the burst duration as the time over which half the burst SNR is accumulated, instead of half the fluence. The SNR frequency width has been smoothly interpolated between the temporally-resolved and -unresolved regimes (Eqs.~\ref{eq:SNR1}, \ref{eq:SNR2}), assuming $t_0 = 0.01$ ms and an instrumental time resolution $\Delta t_{\rm res} = 1$ ms.  The key feature of this plot is the tendency of longer bursts with $\beta \ll 1$ to concentrate their SNR in a narrower bandwidth, while shorter bursts with $\beta \gg 1$ spread their SNR more equally across all frequencies (SNR fractional bandpass 0.5)}
\label{fig:SNR}
\end{figure}

With the connection between the toy model and FRB observables in place, we now move to describe which of the same parameters control the {\it detectability} of a burst and describe how this could shape or bias existing FRB samples. 

\subsection{Signal-to-Noise Ratio}

The CHIME FRB sample is defined largely by a threshold on the burst SNR (\citealt{CHIME+21_catalog}; see their Fig.~24), as are other FRB surveys (e.g., ASKAP, \citealt{Shannon+18}; see \citealt{Keane&Petroff15} for a general discussion).  To predict the measured band-pass of a burst, we must calculate over what frequency range the burst SNR will be accumulated within the framework of the toy model.  

For an instrument sensitivity (system equivalent flux density) $S_{\rm sys}$, the burst SNR is given by the radiometer equation.  Within a single frequency channel of bandwidth $\Delta \nu_{\rm ch}$ and centroid $\nu_i$, the SNR is
\begin{equation}
    {\rm SNR}_i = \frac{\Delta E_{\nu_i}}{\Delta t_{\rm obs,i}\Delta \nu_{\rm ch}} \frac{\sqrt{\Delta \nu_{\rm ch} \Delta t_{\rm obs,i}}}{S_{\rm sys}},
\end{equation}
where $\Delta E_{\nu_i}$ is the burst fluence in the channel and $\Delta t_{\rm obs,i} \sim \max \left( \Delta t_{\rm res} , t_{\rm FRB,i} \right)$ is the observed burst duration, which is the greater of the true duration the burst takes to cross that channel $t_{\rm FRB,i} \approx \Delta \nu_{\rm ch}/\dot{\nu}_c$ (Eq.~\ref{eq:tobs}) and the instrument time-resolution $\Delta t_{\rm res}$.  The total burst SNR is then obtained by summing over all $N = \Delta \nu_{\rm obs}/\Delta \nu_{\rm ch}$ channels:
\begin{align}
    {\rm SNR} 
    &= \frac{1}{\sqrt{N}} \sum_{i=1}^{N} {\rm SNR}_i
    = \frac{1}{S_{\rm sys} \sqrt{\Delta\nu_{\rm obs}}} \sum_{i=1}^{N}\frac{\Delta E_{\nu_i}}{\Delta \nu_{\rm ch}} \Delta t_{{\rm obs},i}^{-1/2} \Delta \nu_{\rm ch}
    \nonumber \\
    &\underset{N \rightarrow \infty, \Delta \nu_{\rm ch} \rightarrow 0}\Rightarrow
    \frac{1}{S_{\rm sys} \sqrt{\Delta\nu_{\rm obs}}} \int_{\nu_{\rm min}}^{\nu_{\rm max}} \frac{dE_\nu}{d\nu} \Delta t_{\rm obs}(\nu)^{-1/2} \, d\nu
    .
\end{align}
For the analytic toy model, this can be expressed more explicitly using $dE_{\nu}/d\nu$ (Eq.~\ref{eq:SED}) to write
\begin{eqnarray}
{\rm SNR} &=& \frac{2}{\sqrt{\pi}}\frac{F_0 t_0}{\beta}\frac{1}{S_{\rm sys} \sqrt{\Delta\nu_{\rm obs}}} \times \nonumber \\
&&\int_{\nu_{\rm min}}^{\nu_{\rm max}'}\left(\frac{\nu}{\nu_0}\right)^{\frac{\alpha-\beta-1}{\beta}}\frac{1}{\sqrt{\Delta t_{\rm obs}(\nu)}} \, d\left(\frac{\nu}{\nu_0}\right).
\end{eqnarray}
For time-resolved bursts with $t_{\rm FRB} > \Delta t_{\rm res}$ we have $\Delta t_{\rm obs} = t_{\rm FRB} = \nu/\dot{\nu}_{\rm c} \sim (t_0/\beta)(\nu/\nu_0)^{-1/\beta}$, and hence
\begin{eqnarray} \label{eq:SNR1}
    {\rm SNR} &=& \frac{2\beta}{2\alpha-1} \frac{1}{S_{\rm sys} \sqrt{\Delta\nu_{\rm obs}}} \frac{2F_0}{\sqrt{\pi}}\left(\frac{t_0}{\beta}\right)^{1/2}\times \nonumber \\ &&
    \left[ \left(\frac{\nu_{\rm max}'}{\nu_0}\right)^{\frac{2\alpha-1}{2\beta}} - \left(\frac{\nu_{\rm min}}{\nu_0}\right)^{\frac{2\alpha-1}{2\beta}} \right], \,\text{Time-Resolved.} \nonumber \\
\end{eqnarray}

On the other hand, if the burst is unresolved throughout its evolution ($t_{\rm FRB} < \Delta t_{\rm res}$; $\Delta t_{\rm obs} = \Delta t_{\rm res}$) then we instead have
\begin{eqnarray} 
    {\rm SNR}&=&  \frac{2}{\sqrt{\pi}} \frac{F_0t_0}{(\alpha-1)} \frac{1}{S_{\rm sys} \sqrt{\Delta\nu_{\rm obs}\Delta t_{\rm res}}}\times \nonumber \\
    && \left[ \left(\frac{\nu_{\rm max}'}{\nu_0}\right)^{\frac{\alpha-1}{\beta}} - \left(\frac{\nu_{\rm min}}{\nu_0}\right)^{\frac{\alpha-1}{\beta}} \right] \nonumber \\
    &=& \frac{E}{S_{\rm sys} \sqrt{\Delta\nu_{\rm obs}\Delta t_{\rm res}}}, \,\,\,\text{Time-Unresolved},
    \label{eq:SNR2}
    \end{eqnarray}
where in the final line we have made use of Eq.~(\ref{eq:fluence}).  Thus, for temporally unresolved bursts, detectability is proportional to the burst fluence.  By contrast, for time-resolved bursts, the SNR is smaller by a factor $\sim (\Delta t_{\rm res}/t_{\rm FRB})^{1/2} < 1$ for typical parameters. 

\subsection{Duration/Bandwidth Dichotomy}

\begin{figure}
    \centering
    \includegraphics[width=0.45\textwidth]{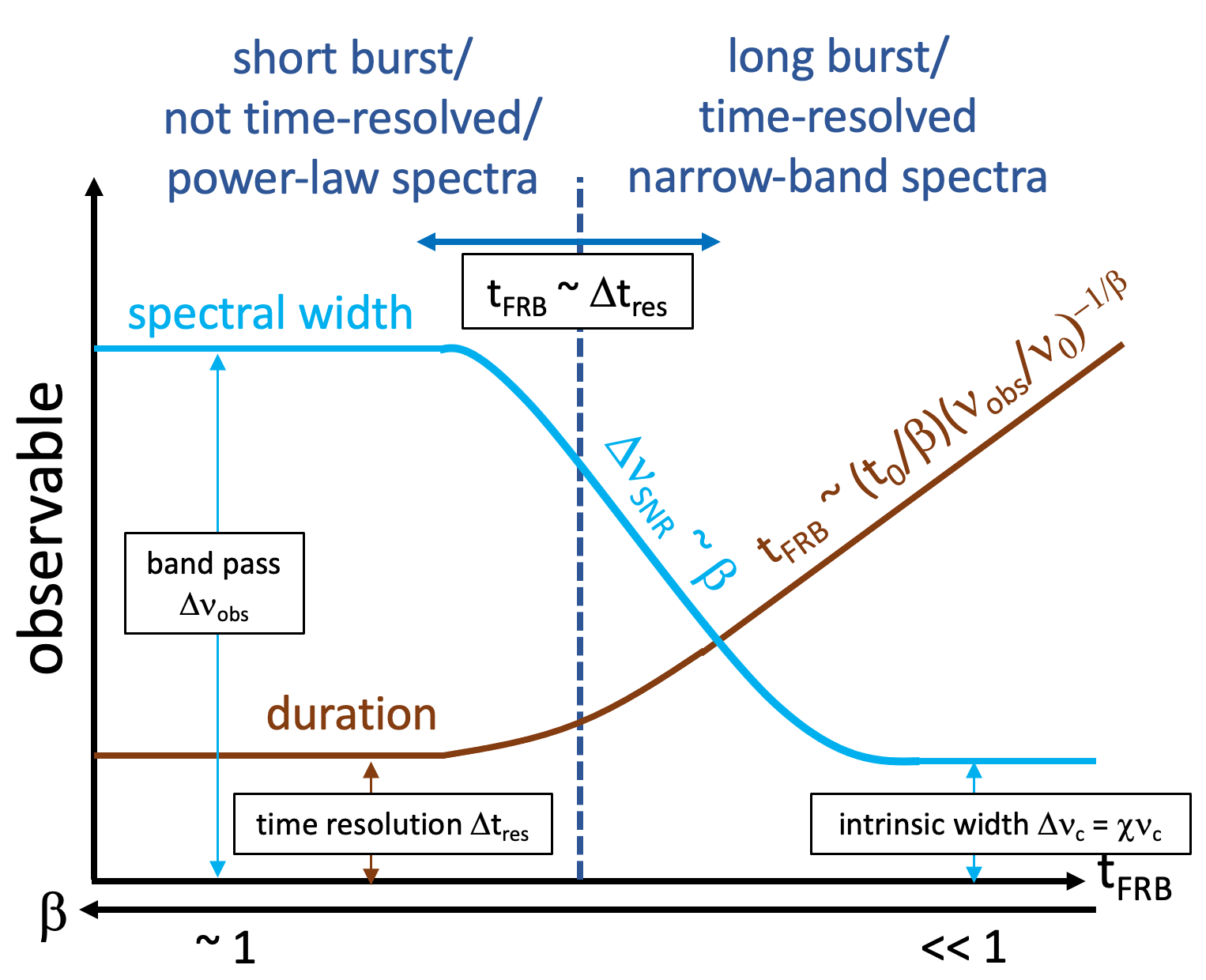}
    \caption{Schematic diagram showing the dependence of observed burst duration $t_{\rm obs}$ (brown line), and time-integrated spectral width (teal line) as a function of the burst duration (or, equivalently, linear drift rate $\beta$) for otherwise fixed values of the spectral resolution $\Delta\nu_{\rm obs}$, time-constant $t_0$, flux normalization $F_0$, and central observing frequency $\nu_{\rm obs}$.  A vertical dot-dashed line shows the temporal resolution of the instrument, which separates unresolved (power-law) spectra of width covering the whole observing band from narrow band spectra down to the intrinsic width $\Delta \nu_{\rm c} \sim \chi \nu$.  }
\label{fig:dichotomy}
\end{figure}

Eqs.~(\ref{eq:SNR1},\ref{eq:SNR2}) reveal several important features.  Firstly, for typical values of $\alpha > 0.5$ ($\alpha > 1$) the burst SNR is peaked at high frequencies in the case of resolved (unresolved) bursts.  Secondly, the ``sharpness'' of this high-frequency bias depends sensitively on the value of $\beta$: low values of $\beta \ll 1$ imply that the burst's cumulative SNR is strongly concentrated at the highest frequencies.  The red dashed contours in Fig.~\ref{fig:SNR} indicate the fractional bandwidth over which $\gtrsim 50\%$ of the total burst SNR is accumulated in the space of $\{\alpha,\beta\}$ for the same burst parameters used in Fig.~\ref{fig:observables}.  

Even given a perfect instrument sensitive to all frequencies emitted by the burst $\in [\nu_0, 0]$, half of the burst SNR would be accumulated in a band-pass $\Delta \nu_{\rm SNR}$ no greater than
\begin{eqnarray}
\frac{\Delta \nu_{\rm SNR}}{\nu_0} &\simeq& 1-\left(\frac{1}{2}\right)^{\frac{2\beta}{2\alpha - 1}} \underset{\beta \ll \alpha - 0.5}\simeq \frac{1.4\beta}{2\alpha-1}, \,\text{ Time-Resolved} \nonumber \\
&\simeq& 1 - \left(\frac{1}{2}\right)^{\frac{\beta}{\alpha-1}} \underset{\beta \ll \alpha - 1}\simeq \frac{0.7\beta}{\alpha-1}, \,\text{ Time-Unresolved} \nonumber \\
\label{eq:nuSNR}
\end{eqnarray}
where we have assumed $\alpha > 1/2$ and $\alpha > 1$, in the resolved and unresolved cases, respectively.  The value of $\Delta \nu_{\rm SNR}$ estimated above is the maximum SNR bandwidth because the derivation of Eq.~(\ref{eq:nuSNR}) implicitly assumes an infinitely narrow intrinsic spectral peak; in reality, the spectral width cannot be smaller than $\Delta \nu_{\rm c}/\nu_{0} \sim \chi$ (e.g., the Gaussian width in Eq.~\ref{eq:nuc}). 

The fact that low values of $\beta$ give rise to narrower SNR-limited spectral widths ($\Delta \nu_{\rm SNR} \propto \beta$) may have important implications for the dichotomy between short- and long-duration FRBs identified by CHIME \citep{Pleunis+21}.  In particular, replacing $\Delta \nu_{\rm obs}$ with $\Delta \nu_{\rm SNR} \sim \beta \nu_0$ (Eq.~\ref{eq:nuSNR} in the $\beta \ll \alpha \sim 1$ limit) in our estimate of $t_{\rm FRB}$ (Eq.~\ref{eq:tobs}), we obtain the approximate relationship:
\be
\beta \propto \frac{\rm ln[\nu_0/\nu_{\rm obs}]}{\rm ln[t_{\rm FRB}/t_0]} \Rightarrow t_{\rm FRB} \propto t_0 \exp\left[\frac{\nu_0}{\Delta \nu_{\rm SNR}}\right].
\ee
This expression demonstrates that the same values of $\beta \ll 1$ that give rise to long bursts ($t_{\rm FRB} \gg t_0$; see also Fig.~\ref{fig:observables}) also correspond to those with narrow-band SNR ($\Delta \nu_{\rm SNR} \ll \nu_0$).  If such a burst were to begin within the instrumental band-pass (i.e., $\nu_0 < \nu_{\rm max}$) it would generate a long-duration burst with a time-integrated spectrum peaked within the band around $\nu \sim \nu_0$. Note that this would remain true even if the burst fluence does not fall off too abruptly with frequency (e.g., $\gamma \approx -1.5$) because the SNR is much more strongly dependent on frequency (${\rm SNR} \propto \nu^{\gamma+1+1/(2\beta)}$). 
Furthermore, if such a low-$\beta$ burst were instead to start well above the instrumental band-pass ($\nu_0 \gg \nu_{\rm max}$), then it would likely go undetected (see discussion below).

Alternatively, the opposite regime, where $\beta$ is not too small ($\beta \sim 1$) imply bursts with shorter durations (Eq.~\ref{eq:tobs}) that accumulate significant SNR across a wider bandpass extending to frequencies $\ll \nu_0$ (i.e., $\Delta \nu_{\rm SNR} \gtrsim \nu_0$). Such bursts, even if they begin well above the instrumental band-pass, can be detected as they sweep down through it and accumulate SNR across a wide range of frequencies, resulting in a short (preferentially unresolved) burst with a broadband (power-law) spectrum across the entire band.

We illustrate the effects just described by means of two figures.  Fig.~\ref{fig:SNR} shows an example of how in the space of $\{\alpha,\beta\}$, bursts with longer duration (solid black contours) also possess narrower SNR bandwidth (red-dashed contours).

\begin{figure*}
    \centering
    \includegraphics[width=0.40\textwidth]{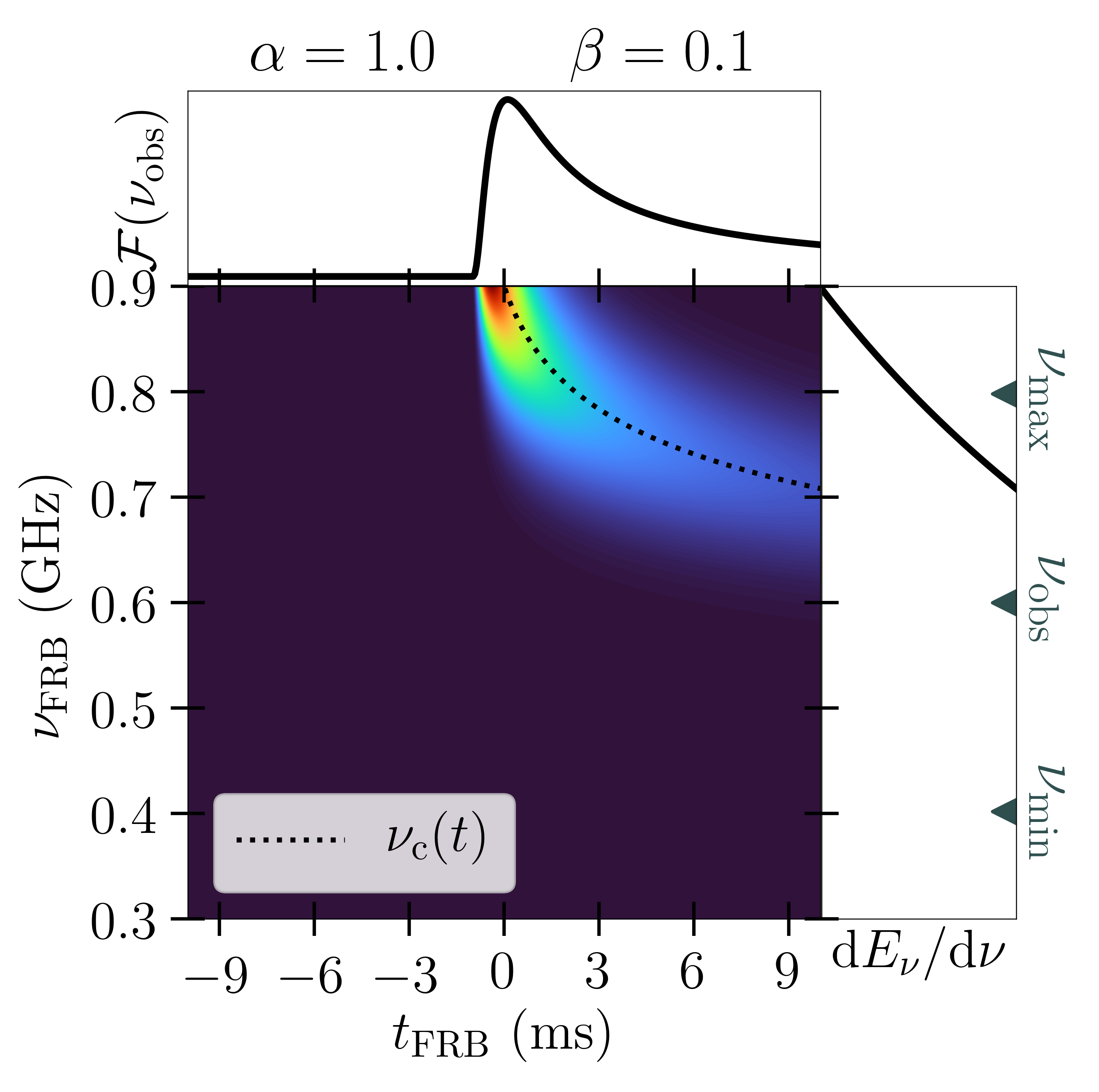}
    \hspace{1.5cm}
    \includegraphics[width=0.475\textwidth]{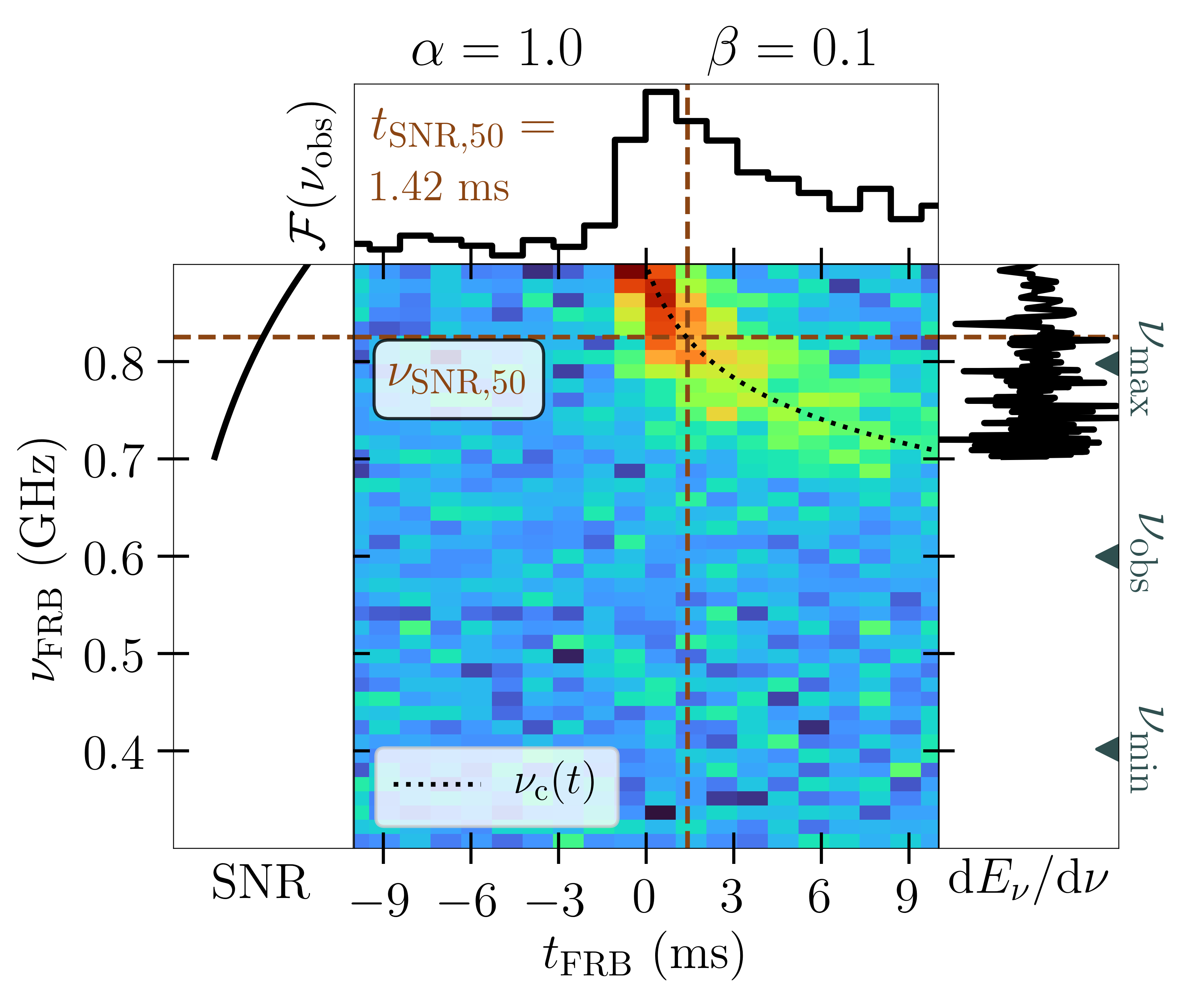}
    \includegraphics[width=0.40\textwidth]{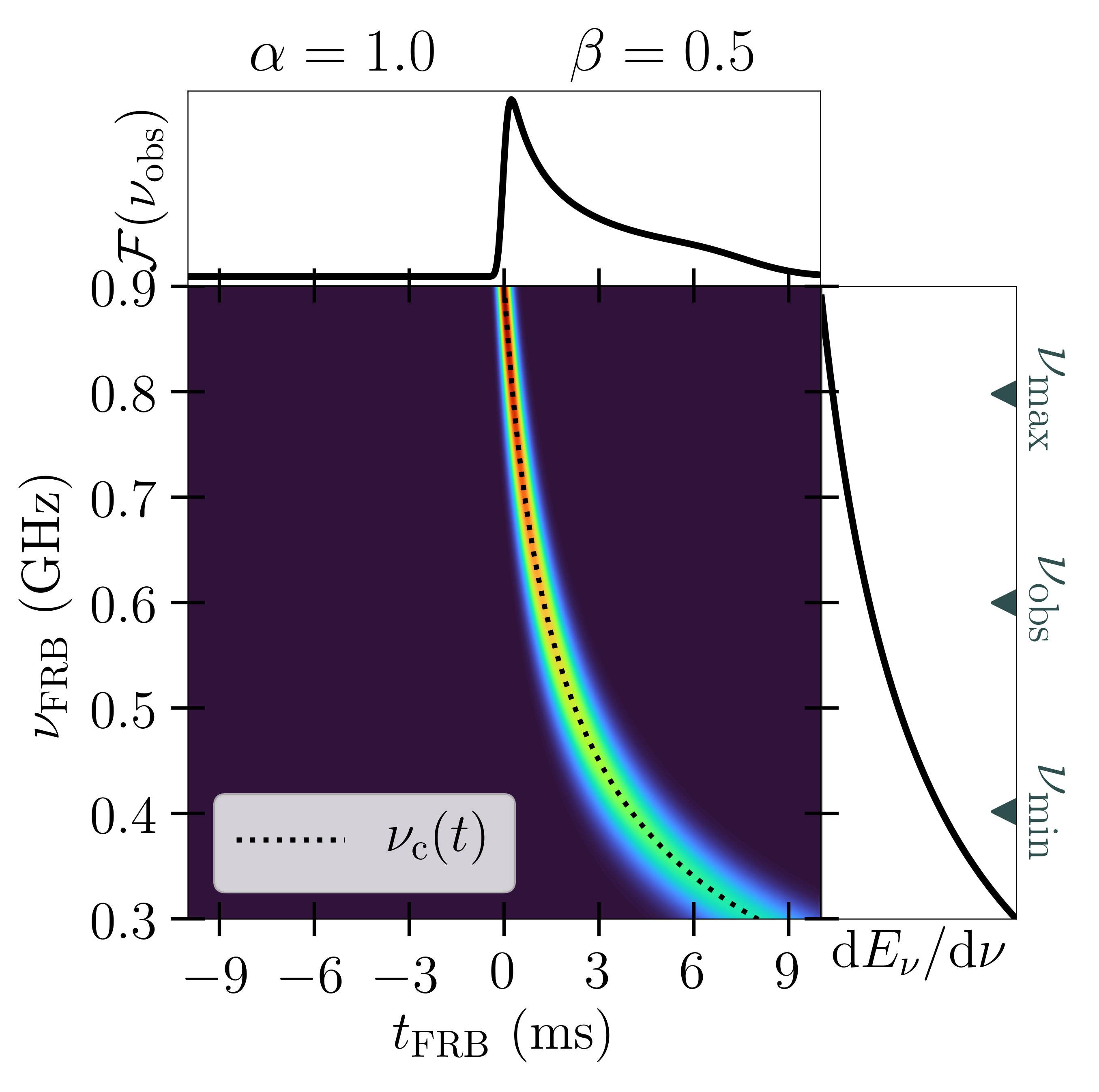}
    \hspace{1.5cm}
    \includegraphics[width=0.475\textwidth]{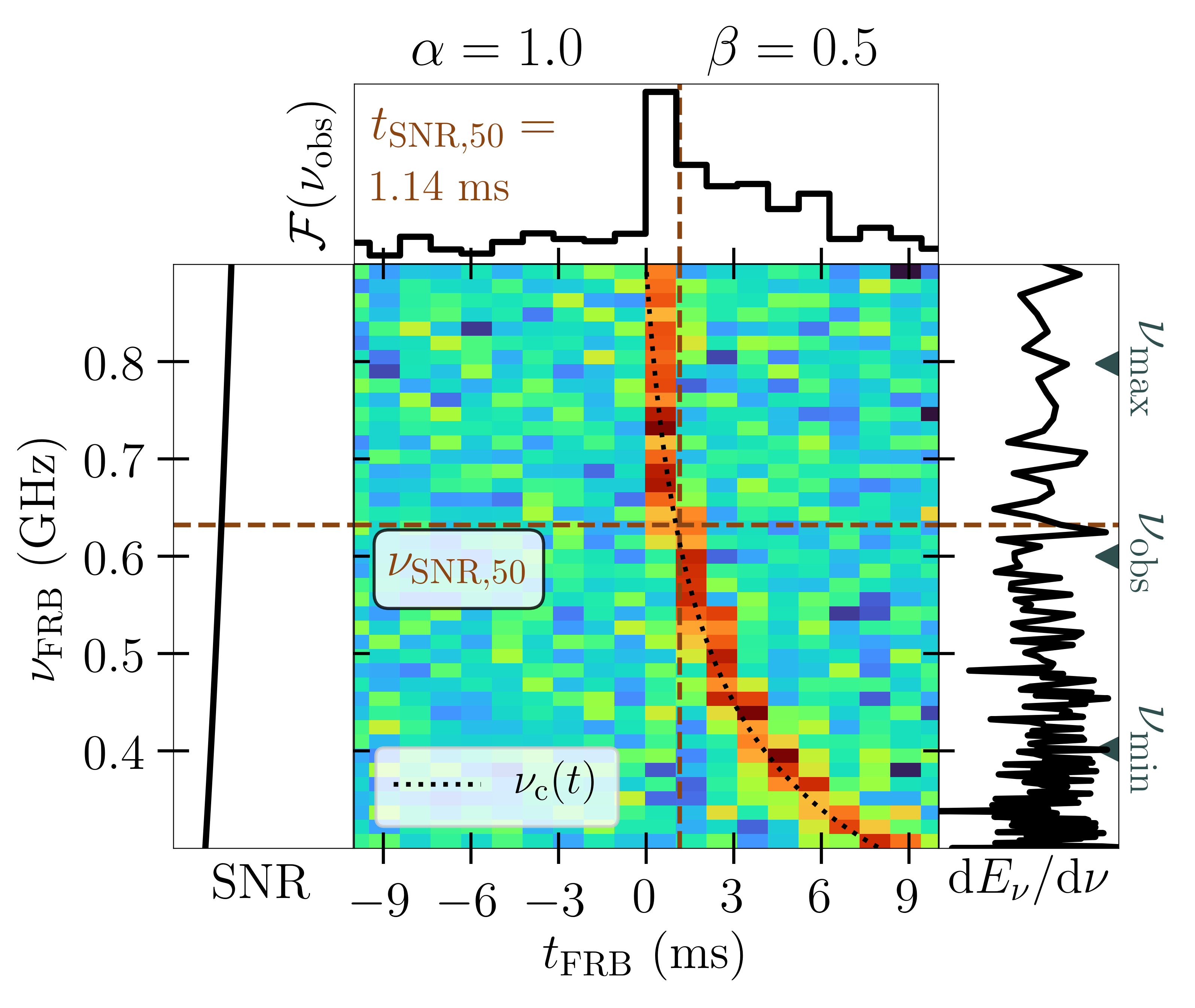}
    \includegraphics[width=0.40\textwidth]{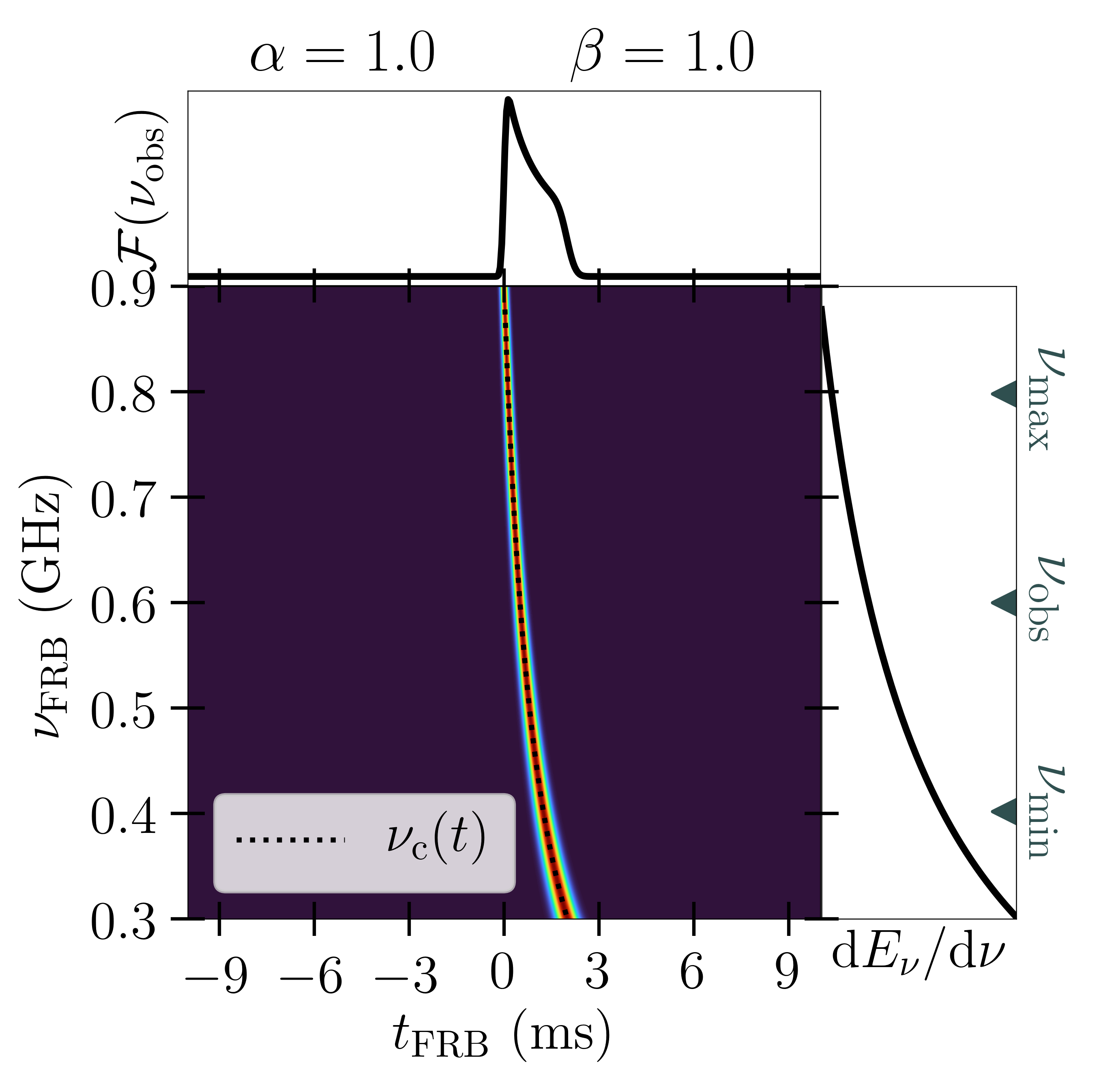}
    \hspace{1.5cm}
    \includegraphics[width=0.475\textwidth]{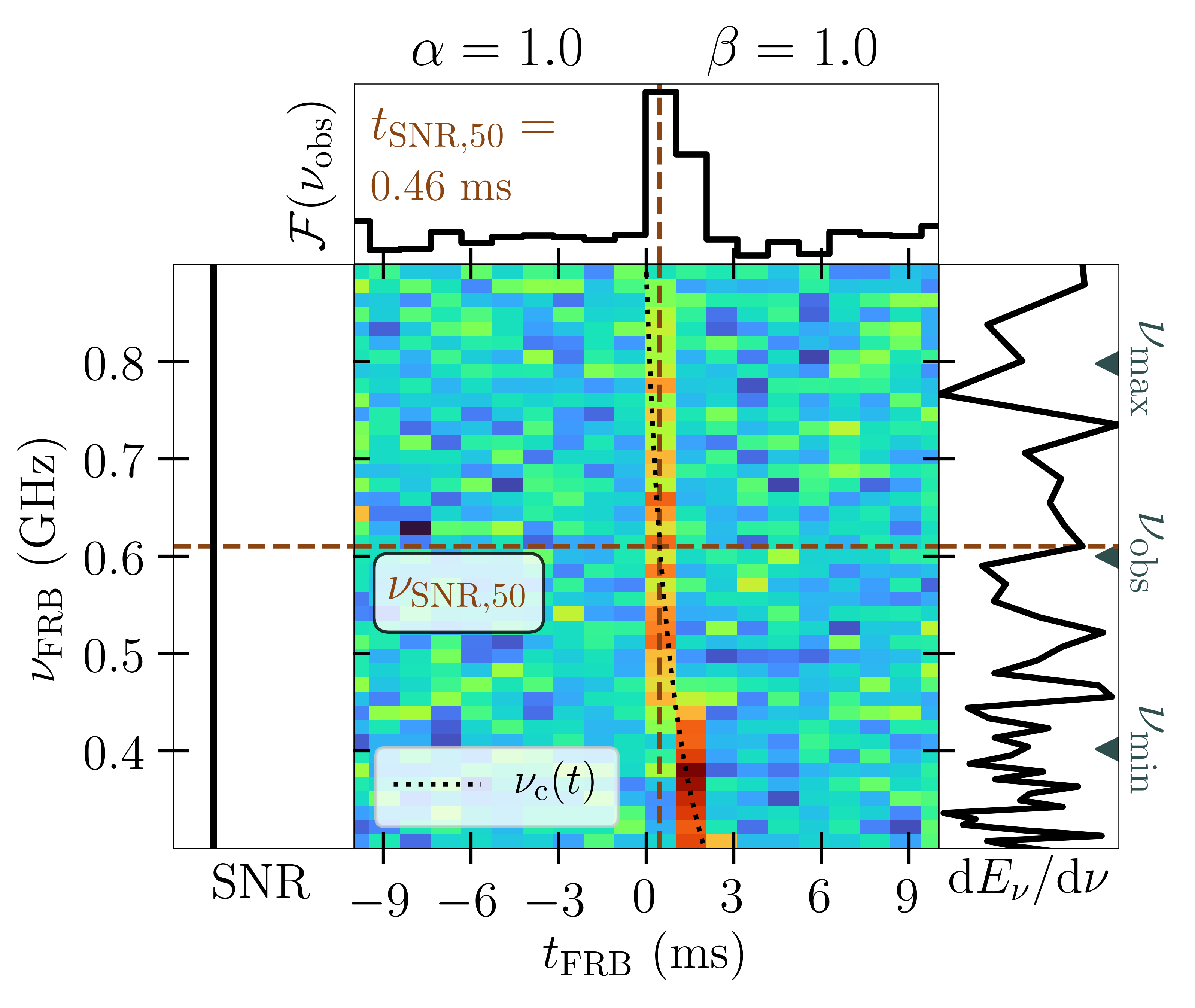}
    \caption{Synthetic time-frequency (``waterfall'') plots calculated for otherwise fixed values of $\alpha = 1$, $\chi = 0.1$, $\nu_0 = 0.9$\,GHz, $t_{\rm 0} = 1$\,ms but varying values of $\beta = 0.1$ (top row), $\beta = 0.5$ (middle row) and $\beta = 1$ (bottom row).  The left column shows the pure signal with zero noise and infinite time/frequency resolution, while in the right column we have introduced time- and frequency-independent white noise and have assumed finite time-resolution $\Delta t_{\rm res} = 1$\,ms. Brown horizontal and vertical dashed lines denote the frequency $\nu_{\rm SNR,50}$ and time $t_{\rm SNR,50}$ over which half the SNR has been accumulated. For each model of FRB, the top sub-panel carries its light curve, and the right sub-panel carries its energy spectrum. The left sub-panel, where applicable, shows the frequency evolution of SNR.}
    \label{fig:noise}
\end{figure*}

Fig.~\ref{fig:noise} shows several example synthetic waterfall plots, with different rows corresponding to bursts with otherwise identical parameters but for three values of $\beta = 0.1, 0.5, 1$.  The left column shows the signal with zero background noise and infinite time resolution, similar to those shown earlier in Fig.~\ref{fig:waterfallschematic}.  The right column shows the same bursts as viewed at a coarser 1\,ms time resolution; the colors here represent flux density averaged over the coarse $t-\nu$ grid. Furthermore in the right column, we also added a time and frequency-independent white noise to the signal in order to mimic the observed data. The SNR thus achieved, at each frequency channel, is presented in the left sub-panel. We quantify the effect of the added noise with the parameters $\nu_{\rm SNR,50}$ (horizontal dashed line) and $t_{\rm SNR,50}$ (vertical dashed line) which denote the frequency and time over which half the SNR has been accumulated. As expected, the SNR evolves rapidly with frequency---accumulating most of the SNR at the top of the band (i.e., large $\nu_{\rm SNR,50}$ and small $t_{\rm SNR,50}$)---for smaller $\beta\sim0.1$, and vice-versa. In accordance with Eqs.~(\ref{eq:SNR1},\ref{eq:SNR2}), we see that for smaller values of $\beta \sim 0.1$, the burst is more readily time-resolved and its spectrum would appear to ``linger'' around a single frequency, while in the $\beta = 1$ case, the burst duration would be hardly resolved, and the spectrum would be broad with an SNR spread evenly over the bandpass. 

In summary, even if all FRBs in nature share a similar distribution of $t_0$, variation of the parameter $\beta$ would act to divide an SNR-limited sample into long time-resolved bursts with narrow spectra ($\beta \ll 1$) and shorter unresolved bursts with broad-band spectra ($\beta \gg 1$).  Fig.~\ref{fig:dichotomy} illustrates this dichotomy schematically.  Confirmation that this effect contributes to the duration-spectral width relationship identified by CHIME \citep{Pleunis+21} would require a more direct modeling of their data within the framework of the toy model.   

We have focused on the effects of varying $\beta$ above, though in principle other parameters of the model could also enter into shaping the burst dichotomy.  Although the burst duration is proportional $t_0$, the frequency width of the burst SNR is insensitive to its value.  Likewise, while the SNR bandwidth is sensitive to $\alpha$ for large $\alpha$, its value is constrained to obey $\alpha \simeq 1-0.5\beta \lesssim 1$ to match the time-integrated spectral index $\gamma \simeq -1.5$ found by \citet{Macquart+19}.

How likely is it that a burst detected in an SNR-limited sample will begin its evolution within the instrumental band-pass (i.e., $\nu_0 < \nu_{\rm max}$), as opposed to above it?  The answer to this question is also related to the band-pass over which the burst SNR is accumulated.  If the SNR is accumulated over only a narrow fraction of the instrument's bandwidth ($\Delta \nu_{\rm SNR} \ll \Delta \nu_{\rm obs}; \beta \ll 1$; top row of Fig.~\ref{fig:noise}), then$-$even assuming that FRB sources in nature sample from a flat distribution of $\nu_0$ values spanning many decades in frequency$-$a burst observed by an SNR-limited survey is just as likely to start at any location within the instrumental band-pass ($\nu_{\rm min} < \nu_0 < \nu_{\rm max}$) as it would marginally above it ($\nu_{\rm max} < \nu_0 < \nu_{\rm max} + \Delta \nu_{\rm obs}/2$).  This is again consistent with FRB observations that show longer bursts with narrower time-averaged spectra do not always start at the top of the band, while the shorter bursts more frequently spread their SNR across a wider range of frequencies (and thus began their downwards frequency evolution starting above the instrumental band-pass).  

\section{Conclusion}
\label{sec:conclusions}

We have presented a toy model for the time-frequency structure of FRBs.  The key feature is a narrowly peaked SED whose peak frequency and luminosity scale as power-laws with respect to a singular point in time (corresponding, e.g., to the onset of energy release from the central engine).  The model can naturally accommodate many observed FRB properties, including: narrow spectral widths of time-resolved bursts and broad power-law spectra of unresolved bursts; downward frequency evolution of burst structure across the burst duration; tendency for shorter intrinsic burst durations at higher radio frequencies; and larger temporal separation between separate burst components for bursts of total longer bursts (Fig.~\ref{fig:components}).  Our model is independent of the FRB emission mechanism, though decelerating relativistic shocks (e.g., \citealt{Metzger+19}) or radius-frequency mapping in a neutron star magnetosphere (e.g., \citealt{Lyutikov2020}) offer possible physical realizations.

We have shown that burst-to-burst variation in a single parameter of the model$-$the power-law index of the frequency drift rate $\beta-$naturally generates a dichotomy in burst properties: narrow slowly drifting spectral peaks for temporally resolved longer bursts, and broader power-law spectra for unresolved shorter bursts, similar to the dichotomy in the CHIME/FRB sample \citep{CHIME+21_catalog}.  This separation arises because bursts accumulate their SNR over a narrower frequency bandpass when the intrinsic SED drifts downwards in time more slowly.  In cases when $\beta$ is sufficiently large, all the burst SNR is accumulated at roughly the initial frequency $\nu_0$ and the burst may appear at a single frequency in the middle of the observing band that does not evolve in time (e.g., FRB 20190515D, a narrow-band Gaussian burst centered around 550 MHz; see Fig. 3 of \citealt{Pleunis+21}).  

Assuming the FRB duration/spectral-width dichotomy can be attributed to a continuous variation of burst parameters (e.g., $\beta$), this does itself not address why bursts from repeating and non-repeating FRB sources would preferentially exhibit these different parameters (e.g., smaller and larger values of $\beta$, respectively).  We cannot deduce any compelling reason why bursts with narrower spectra but otherwise similar burst properties would preferentially have multiple bursts detected and therefore be categorized as repeaters.  This suggests the required diversity in $\beta$ may lie with intrinsic differences in the central engines or environments between repeaters and non-repeaters.

In scenarios for FRB emission arising from a relativistic shock propagating into a sub-relativistically expanding upstream medium \citep{Metzger+19}, the value of $\beta$ is connected to the density profile of the upstream medium $n \propto r^{-k}$.  For example, in the ``bare shock ($t \gtrsim \delta t$)'' scenario reviewed in Appendix \ref{app:shock} (Table \ref{tab:shock}), one predicts
\be
k = \frac{8\beta - 3}{2\beta} ,  
\ee
such that $\beta \in [0.3,1]$ maps into $k \in [-1,2.5]$.  All else being equal, longer/narrow-band bursts (low $\beta$) would arise from shocks entering a flat or rising density profile ($k \lesssim 0$), while shorter/broader-band bursts (high $\beta$) would arise from shocks entering a wind type medium ($k \gtrsim 2$).  If the upstream medium is produced by the ejecta shell of a previous burst or bursts (e.g., \citealt{Metzger+19}), then it would not be surprising for more frequently repeating FRB sources to be surrounded by a different external medium than those from sources which repeat less frequently.  Alternatively, in shock scenarios in which the upstream medium is that of a relativistic wind (e.g., of a magnetar - \citealt{Beloborodov20} or a binary neutron star merger - \citealt{Sridhar+21}), the typical values of $\beta$ may be larger $\sim 2-3$ (Eq.~\ref{eq:BNS}), and hence bursts produced by this mechanism may preferentially be of shorter-duration (for otherwise similar physical size of the central engine, $\propto t_0$).  

Two of the best studied repeating FRB sources, FRB180916 and FRB 121102, have exhibited periodicity in the burst arrival times of 16 and 160 days, respectively, \citep{CHIME+20b,Rajwade+20}, in which windows of FRB activity are separated by long `dead' periods with little or no burst activity.  This periodic behavior may be related to rotation or precession of the central engine (e.g., \citealt{Beniamini+19,Lyutikov_20,Levin+20,Sridhar+21b}).  If the value of $\beta$ is related is a systematic way to the burst environment, we could expect a systematic variation in $\beta$ (and hence of the burst duration or spectral width) across the active phase window.

\acknowledgements  We thank Shami Chatterjee and Ziggy Pleunis for helpful conversations.  BDM acknowledges support from the National Science Foundation (grant \# GG016244). The research of PB was funded by the Gordon and Betty Moore Foundation through Grant GBMF5076.
BM is supported by NASA through the NASA Hubble Fellowship grant \#HST-HF2-51412.001-A awarded by the Space Telescope Science Institute, which is operated by the Association of Universities for Research in Astronomy, Inc., for NASA, under contract NAS5-26555. LS acknowledges support from the Cottrell Scholars Award and NASA 80NSSC18K1104.

\appendix

\section{Toy Model Parameters for Shock Emission Scenario}
\label{app:shock}

\begin{table}[h!]
  \begin{center}
    \caption{Toy model power-law indices for maser emission from decelerating relativistic shock propagation into a quasi-stationary external medium with power-law radial density profile $n \propto r^{-k}$.}
    \label{tab:table1}
    \begin{tabular}{c|c|c|c|c} 
      Case & - & $\beta$ &  $\alpha$ & $\gamma = \frac{\alpha-\beta-1}{\beta} $ \\
      \hline 
\hline 
      Bare Shock & $t  \lesssim \delta t$ & $\frac{(k+2)}{2(4-k)} \underset{k=0}= 0.25$  & 0 & $\frac{(k-10)}{(k+2)} \underset{k=0}=-5$ \\
      ($\nu_{\rm c}, \Delta \nu_{\rm c} \propto \nu_{\rm pk}$)  & $t \gtrsim \delta t$ & $\frac{3}{2(4-k)} \underset{k=0}= 0.38$  & 1 & -1  \\
\hline
ICS Absorbed & $t  \lesssim \delta t$ & $\frac{2+7k}{8(4-k)} \underset{k=0}=0.06$  & $\frac{6-3k}{8(4-k)} \underset{k=0}= 0.19$ & $-\frac{28-4k}{2+7k}\underset{k=0}=-14$ \\
       ($\nu_{\rm c}, \Delta \nu_{\rm c} \propto \nu_{\rm max}$)  & $t  \gtrsim \delta t$ & $\frac{2k+7}{4(8-2k)} \underset{k=0}=0.22$  & $\frac{(37-10k)}{8(4-2k)} \underset{k=0}=1.16$ & ...$\underset{k=0}=-0.29$ \\
    \end{tabular}
\label{tab:shock}
  \end{center}
\end{table}

This section briefly reviews the predicted values of the toy model parameters $\{\alpha,\beta\}$ within a model in which FRB emission is generated by synchrotron maser emission from a relativistic magnetized shock (e.g., \citealt{Lyubarsky14,Beloborodov17}) in the case of a stationary upstream medium or one which is expanding sub-relativistically (\citealt{Metzger+19}).  

Consider an impulsive injection of energy over time $\delta t$ which generates a shock propagating into an stationary external medium with a radial density profile $n \propto r^{-k}$.  The shock will decelerate in a self-similar manner (\citealt{Blandford_McKee_76}), such that the Lorentz factor, density, and kinetic luminosity of the shock at observer time $t$ after the injection, obey:
\begin{equation}\Gamma  \propto \begin{cases}
t^{\frac{(k-2)}{2(4-k)}} \underset{k=0}\approx t^{-1/4}, & t < \delta t  \\
t^{\frac{(k-3)}{2(4-k)}} \underset{k=0}\approx t^{-3/8}, & t > \delta t,
\end{cases}
\label{eq:gammashock}
\end{equation}
\begin{equation} n   \propto \begin{cases}
t^{\frac{-2k}{(4-k)}} \underset{k=0}\approx t^{0}, & t < \delta t  \\
t^{\frac{-k}{(4-k)}} \underset{k=0}\approx t^{0}, & t > \delta t,
\end{cases}
\label{eq:nshock}
\end{equation}
\begin{equation} L_{\rm sh}   \propto \begin{cases}
t^{0}, & t < \delta t  \\
 t^{-1}, & t > \delta t,
\end{cases}
\label{eq:Lshock}
\end{equation}
where separate scalings are given for $t < \delta t$ and $t > \delta t$.\\

For a fixed magnetization of the upstream medium, the synchtrotron maser emission will peak at a frequency (e.g., \citealt{Plotnikov&Sironi19}),
\begin{equation} \nu_{\rm pk} \simeq \Gamma \nu_{\rm pk}' \propto \begin{cases}
t^{-\frac{(k+2)}{2(4-k)}} \underset{k=0}\approx t^{-1/4}, & t < \delta t  \\
t^{-\frac{3}{2(4-k)}} \underset{k=0}\approx t^{-3/8}, & t > \delta t,
\end{cases}
\label{eq:nupk}
\end{equation}

However, due to induced Compton scattering (ICS), only frequency $\nu > \nu_{\rm max}$ is able to escape from the upstream, where \citep{Metzger+19,Margalit+20}
\begin{equation} 
\nu_{\rm max}  \propto \begin{cases} t^{-\frac{2+7k}{4(8-2k)}} \underset{k=0}\approx t^{-1/16}, & t < \delta t  \\
t^{-\frac{2k+7}{4(8-2k)}} \underset{k=0}\approx t^{-7/32}, & t > \delta t,
\end{cases}
\label{eq:numax}
\end{equation}

The total power of the synchrotron maser emission (most of which emerges around $\nu_{\rm pk}$) scales with the power of the shock, $L_{\rm sh}$.  Assuming the synchrotron maser emission possesses a power-law spectrum $\nu L_{\nu} \propto \nu^{3-\kappa}$ for $\nu \gtrsim \nu_{\rm pk}$, where $\kappa \approx 4$ is estimated from the results of particle-in-cell shock simulations (e.g., \citealt{Plotnikov&Sironi19}), then the observed radio luminosity (at the marginally-thin frequency $\nu_{\rm max}$) obeys  (e.g., \citealt{Margalit+20})
\be
L_{\rm FRB} \propto L_{\rm sh}|_{\nu_{\rm pk}} \left(\frac{\nu_{\rm max}}{\nu_{\rm pk}}\right)^{3-\kappa} \underset{\kappa \approx 4}\propto L_{\rm sh}\frac{\nu_{\rm pk}}{\nu_{\rm max}} 
\ee
and hence in the case of ICS-absorbed emission we have
\begin{equation} 
L_{\rm FRB} \propto \begin{cases} t^{-\frac{6-3k}{8(4-k)}} \underset{k=0}\approx t^{-3/16}, & t < \delta t  \\ t^{-\frac{37-10k}{8(4-k)}} \underset{k=0}\approx t^{-37/32}, & t > \delta t,
\end{cases}
\label{eq:LFRB}
\end{equation}

There are two possibilities to consider, as summarized in Table \ref{tab:shock}.  If the intrinsic narrow spectra of the FRB emission results from a combination of a decaying spectrum of the maser emission towards higher frequencies, and attenuation due to induced Compton scattering at low frequencies (`ICS Absorbed' model; \citealt{Metzger+19}), then both $\nu_{\rm c}$ and $\Delta \nu_{\rm c}$ should scale $\propto \nu_{\rm max}$ (Eq.~\ref{eq:numax}).  On the other hand, if the observed spectral peak is instead a narrow feature intrinsic to the maser emission (e.g., \citealt{Babul&Sironi20}) emitted into an otherwise transparent upstream (`Bare Shock' model), then we instead expect $\Delta \nu_{\rm c}, \nu_{\rm c} \propto \nu_{\rm pk}$ (Eq.~\ref{eq:nupk}).  

As far as the luminosity normalization (Eq.~\ref{eq:integrated}), the simplest version of the ICS Absorbed model predicts $\int F_{\nu}d\nu \propto L_{\rm sh}(\nu_{\rm pk}/\nu_{\rm max})$ (Eq.~\ref{eq:LFRB}).  By contrast, in the Bare Shock model one instead will have $\int F_{\nu}d\nu \propto L_{\rm sh} \propto t^{-1}$ for $t \gtrsim \delta t$ (Eq.~\ref{eq:Lshock}), i.e. $\alpha = 1$ independent of $k$.  The luminosity decay is shallower ($\alpha \approx 0$) at early times when the reverse shock is still crossing through the ejecta shell ($t < \delta t$).


\begin{thebibliography}{}
\expandafter\ifx\csname natexlab\endcsname\relax\def\natexlab#1{#1}\fi
\providecommand{\url}[1]{\href{#1}{#1}}
\providecommand{\dodoi}[1]{doi:~\href{http://doi.org/#1}{\nolinkurl{#1}}}
\providecommand{\doeprint}[1]{\href{http://ascl.net/#1}{\nolinkurl{http://ascl.net/#1}}}
\providecommand{\doarXiv}[1]{\href{https://arxiv.org/abs/#1}{\nolinkurl{https://arxiv.org/abs/#1}}}

\bibitem[{{Babul} \& {Sironi}(2020)}]{Babul&Sironi20}
{Babul}, A.-N., \& {Sironi}, L. 2020, \mnras, 499, 2884,
  \dodoi{10.1093/mnras/staa2612}

\bibitem[{{Beloborodov}(2017)}]{Beloborodov17}
{Beloborodov}, A.~M. 2017, \apjl, 843, L26, \dodoi{10.3847/2041-8213/aa78f3}

\bibitem[{{Beloborodov}(2019)}]{Beloborodov19}
---. 2019, arXiv e-prints, arXiv:1908.07743.
\newblock \doarXiv{1908.07743}

\bibitem[{{Beloborodov}(2020)}]{Beloborodov20}
---. 2020, \apj, 896, 142, \dodoi{10.3847/1538-4357/ab83eb}

\bibitem[{{Beniamini} \& {Kumar}(2020)}]{Beniamini&Kumar20}
{Beniamini}, P., \& {Kumar}, P. 2020, \mnras, 498, 651,
  \dodoi{10.1093/mnras/staa2489}

\bibitem[{{Beniamini} {et~al.}(2021){Beniamini}, {Kumar}, \&
  {Narayan}}]{BKN2021}
{Beniamini}, P., {Kumar}, P., \& {Narayan}, R. 2021, arXiv e-prints,
  arXiv:2110.00028.
\newblock \doarXiv{2110.00028}

\bibitem[{{Beniamini} {et~al.}(2019){Beniamini}, {Petropoulou}, {Barniol
  Duran}, \& {Giannios}}]{Beniamini+19}
{Beniamini}, P., {Petropoulou}, M., {Barniol Duran}, R., \& {Giannios}, D.
  2019, \mnras, 483, 840, \dodoi{10.1093/mnras/sty3093}

\bibitem[{{Blandford} \& {McKee}(1976)}]{Blandford_McKee_76}
{Blandford}, R.~D., \& {McKee}, C.~F. 1976, Physics of Fluids, 19, 1130,
  \dodoi{10.1063/1.861619}

\bibitem[{{Caleb} {et~al.}(2020){Caleb}, {Stappers}, {Abbott}, {Barr},
  {Bezuidenhout}, {Buchner}, {Burgay}, {et~al.}}]{Caleb_20}
{Caleb}, M., {Stappers}, B.~W., {Abbott}, T.~D., {et~al.} 2020, \mnras, 496,
  4565, \dodoi{10.1093/mnras/staa1791}

\bibitem[{{Chamma} {et~al.}(2020){Chamma}, {Rajabi}, {Wyenberg}, {Mathews}, \&
  {Houde}}]{Chamma+21}
{Chamma}, M.~A., {Rajabi}, F., {Wyenberg}, C.~M., {Mathews}, A., \& {Houde}, M.
  2020, arXiv e-prints, arXiv:2010.14041.
\newblock \doarXiv{2010.14041}

\bibitem[{{Chime/Frb Collaboration} {et~al.}(2020){Chime/Frb Collaboration},
  {Amiri}, {Andersen}, {et~al.}}]{CHIME+20b}
{Chime/Frb Collaboration}, {Amiri}, M., {Andersen}, B.~C., {et~al.} 2020, \nat,
  582, 351, \dodoi{10.1038/s41586-020-2398-2}

\bibitem[{{CHIME/FRB Collaboration} {et~al.}(2019){CHIME/FRB Collaboration},
  {Andersen}, {Bandura}, {et~al.}}]{CHIME+19repeaters}
{CHIME/FRB Collaboration}, {Andersen}, B.~C., {Bandura}, K., {et~al.} 2019,
  \apjl, 885, L24, \dodoi{10.3847/2041-8213/ab4a80}

\bibitem[{{Cho} {et~al.}(2020){Cho}, {Macquart}, {Shannon}, {Deller},
  {Morrison}, {Ekers}, {Bannister}, {Farah}, {Qiu}, {Sammons}, {Bailes},
  {Bhandari}, {Day}, {James}, {Phillips}, {Prochaska}, \& {Tuthill}}]{Cho+20}
{Cho}, H., {Macquart}, J.-P., {Shannon}, R.~M., {et~al.} 2020, \apjl, 891, L38,
  \dodoi{10.3847/2041-8213/ab7824}

\bibitem[{{Cordes} \& {Chatterjee}(2019)}]{Cordes&Chatterjee19}
{Cordes}, J.~M., \& {Chatterjee}, S. 2019, \araa, 57, 417,
  \dodoi{10.1146/annurev-astro-091918-104501}

\bibitem[{{Cordes} {et~al.}(2017){Cordes}, {Wasserman}, {Hessels}, {Lazio},
  {Chatterjee}, \& {Wharton}}]{Cordes+17}
{Cordes}, J.~M., {Wasserman}, I., {Hessels}, J.~W.~T., {et~al.} 2017, \apj,
  842, 35, \dodoi{10.3847/1538-4357/aa74da}

\bibitem[{{Day} {et~al.}(2020){Day}, {Deller}, {Shannon}, {Qiu}, {Bannister},
  {Bhandari}, {Ekers}, {Flynn}, {James}, {Macquart}, {Mahony}, {Phillips}, \&
  {Xavier Prochaska}}]{Day+20}
{Day}, C.~K., {Deller}, A.~T., {Shannon}, R.~M., {et~al.} 2020, \mnras, 497,
  3335, \dodoi{10.1093/mnras/staa2138}

\bibitem[{{Falcke} \& {Rezzolla}(2014)}]{Falcke&Rezzolla14}
{Falcke}, H., \& {Rezzolla}, L. 2014, \aap, 562, A137,
  \dodoi{10.1051/0004-6361/201321996}

\bibitem[{{Farah} {et~al.}(2018){Farah}, {Flynn}, {Bailes}, {Jameson},
  {et~al.}}]{Farah+18}
{Farah}, W., {Flynn}, C., {Bailes}, M., {Jameson}, A., {et~al.} 2018, \mnras,
  478, 1209, \dodoi{10.1093/mnras/sty1122}

\bibitem[{{Fuller} \& {Ott}(2015)}]{Fuller&Ott15}
{Fuller}, J., \& {Ott}, C.~D. 2015, \mnras, 450, L71,
  \dodoi{10.1093/mnrasl/slv049}

\bibitem[{{Gajjar} {et~al.}(2018){Gajjar}, {Siemion}, {Price}, {Law},
  {Michilli}, {Hessels}, {Chatterjee}, {Archibald}, {Bower}, {Brinkman},
  {Burke-Spolaor}, {Cordes}, {Croft}, {Enriquez}, {Foster}, {Gizani},
  {Hellbourg}, {Isaacson}, {Kaspi}, {Lazio}, {Lebofsky}, {Lynch}, {MacMahon},
  {McLaughlin}, {Ransom}, {Scholz}, {Seymour}, {Spitler}, {Tendulkar},
  {Werthimer}, \& {Zhang}}]{Gajjar_18}
{Gajjar}, V., {Siemion}, A.~P.~V., {Price}, D.~C., {et~al.} 2018, \apj, 863, 2,
  \dodoi{10.3847/1538-4357/aad005}

\bibitem[{{Hashimoto} {et~al.}(2020){Hashimoto}, {Goto}, {Wang}, {Kim}, {Ho},
  {On}, {Lu}, \& {Santos}}]{Hashimoto+20}
{Hashimoto}, T., {Goto}, T., {Wang}, T.-W., {et~al.} 2020, \mnras, 494, 2886,
  \dodoi{10.1093/mnras/staa895}

\bibitem[{{Hessels} {et~al.}(2019){Hessels}, {Spitler}, {Seymour},
  {et~al.}}]{Hessels+19}
{Hessels}, J.~W.~T., {Spitler}, L.~G., {Seymour}, A.~D., {et~al.} 2019, \apjl,
  876, L23, \dodoi{10.3847/2041-8213/ab13ae}

\bibitem[{{Ioka} \& {Zhang}(2020)}]{Ioka&Zhang20}
{Ioka}, K., \& {Zhang}, B. 2020, \apjl, 893, L26,
  \dodoi{10.3847/2041-8213/ab83fb}

\bibitem[{{Josephy} {et~al.}(2019){Josephy}, {Chawla}, {Fonseca},
  {et~al.}}]{Josephy+19}
{Josephy}, A., {Chawla}, P., {Fonseca}, E., {et~al.} 2019, \apjl, 882, L18,
  \dodoi{10.3847/2041-8213/ab2c00}

\bibitem[{{Keane} \& {Petroff}(2015)}]{Keane&Petroff15}
{Keane}, E.~F., \& {Petroff}, E. 2015, \mnras, 447, 2852,
  \dodoi{10.1093/mnras/stu2650}

\bibitem[{{Kumar} \& {Bo{\v{s}}njak}(2020)}]{Kumar&Bosnjak20}
{Kumar}, P., \& {Bo{\v{s}}njak}, {\v{Z}}. 2020, \mnras, 494, 2385,
  \dodoi{10.1093/mnras/staa774}

\bibitem[{{Kumar} {et~al.}(2017){Kumar}, {Lu}, \& {Bhattacharya}}]{Kumar+17}
{Kumar}, P., {Lu}, W., \& {Bhattacharya}, M. 2017, \mnras, 468, 2726,
  \dodoi{10.1093/mnras/stx665}

\bibitem[{{Kumar} {et~al.}(2021){Kumar}, {Shannon}, {Flynn}, {Os{\l}owski},
  {Bhandari}, {Day}, {Deller}, {Farah}, {Kaczmarek}, {Kerr}, {Phillips},
  {Price}, {Qiu}, \& {Thyagarajan}}]{Kumar+21}
{Kumar}, P., {Shannon}, R.~M., {Flynn}, C., {et~al.} 2021, \mnras, 500, 2525,
  \dodoi{10.1093/mnras/staa3436}

\bibitem[{{Levin} {et~al.}(2020){Levin}, {Beloborodov}, \&
  {Bransgrove}}]{Levin+20}
{Levin}, Y., {Beloborodov}, A.~M., \& {Bransgrove}, A. 2020, arXiv e-prints,
  arXiv:2002.04595.
\newblock \doarXiv{2002.04595}

\bibitem[{{Li} {et~al.}(2021){Li}, {Dong}, {Zhang}, \& {Li}}]{Li+21}
{Li}, X.~J., {Dong}, X.~F., {Zhang}, Z.~B., \& {Li}, D. 2021, arXiv e-prints,
  arXiv:2110.07227.
\newblock \doarXiv{2110.07227}

\bibitem[{{Lu} {et~al.}(2021){Lu}, {Beniamini}, \& {Kumar}}]{LBK2021}
{Lu}, W., {Beniamini}, P., \& {Kumar}, P. 2021, arXiv e-prints,
  arXiv:2107.04059.
\newblock \doarXiv{2107.04059}

\bibitem[{{Lu} {et~al.}(2020){Lu}, {Piro}, \& {Waxman}}]{Lu+20}
{Lu}, W., {Piro}, A.~L., \& {Waxman}, E. 2020, arXiv e-prints,
  arXiv:2003.12581.
\newblock \doarXiv{2003.12581}

\bibitem[{{Lyubarsky}(2014)}]{Lyubarsky14}
{Lyubarsky}, Y. 2014, \mnras, 442, L9, \dodoi{10.1093/mnrasl/slu046}

\bibitem[{{Lyubarsky}(2019)}]{Lyubarsky19}
---. 2019, \mnras, 483, 1731, \dodoi{10.1093/mnras/sty3233}

\bibitem[{{Lyubarsky}(2020)}]{Lyubarsky20}
---. 2020, \apj, 897, 1, \dodoi{10.3847/1538-4357/ab97b5}

\bibitem[{{Lyutikov}(2020)}]{Lyutikov2020}
{Lyutikov}, M. 2020, \apj, 889, 135, \dodoi{10.3847/1538-4357/ab55de}

\bibitem[{{Lyutikov}(2021)}]{Lyutikov21}
---. 2021, arXiv e-prints, arXiv:2102.07010.
\newblock \doarXiv{2102.07010}

\bibitem[{{Lyutikov} {et~al.}(2020){Lyutikov}, {Barkov}, \&
  {Giannios}}]{Lyutikov_20}
{Lyutikov}, M., {Barkov}, M.~V., \& {Giannios}, D. 2020, \apjl, 893, L39,
  \dodoi{10.3847/2041-8213/ab87a4}

\bibitem[{{Macquart} {et~al.}(2019){Macquart}, {Shannon}, {Bannister}, {James},
  {Ekers}, \& {Bunton}}]{Macquart+19}
{Macquart}, J.~P., {Shannon}, R.~M., {Bannister}, K.~W., {et~al.} 2019, \apjl,
  872, L19, \dodoi{10.3847/2041-8213/ab03d6}

\bibitem[{{Majid} {et~al.}(2020){Majid}, {Pearlman}, {Nimmo}, {Hessels},
  {Prince}, {Naudet}, {Kocz}, \& {Horiuchi}}]{Majid+20}
{Majid}, W.~A., {Pearlman}, A.~B., {Nimmo}, K., {et~al.} 2020, \apjl, 897, L4,
  \dodoi{10.3847/2041-8213/ab9a4a}

\bibitem[{{Margalit} {et~al.}(2019){Margalit}, {Berger}, \&
  {Metzger}}]{Margalit+19}
{Margalit}, B., {Berger}, E., \& {Metzger}, B.~D. 2019, \apj, 886, 110,
  \dodoi{10.3847/1538-4357/ab4c31}

\bibitem[{{Margalit} {et~al.}(2020){Margalit}, {Metzger}, \&
  {Sironi}}]{Margalit+20}
{Margalit}, B., {Metzger}, B.~D., \& {Sironi}, L. 2020, \mnras,
  \dodoi{10.1093/mnras/staa1036}

\bibitem[{{Metzger} {et~al.}(2019){Metzger}, {Margalit}, \&
  {Sironi}}]{Metzger+19}
{Metzger}, B.~D., {Margalit}, B., \& {Sironi}, L. 2019, \mnras, 485, 4091,
  \dodoi{10.1093/mnras/stz700}

\bibitem[{{Michilli} {et~al.}(2018){Michilli}, {Seymour}, {Hessels},
  {et~al.}}]{Michilli+18}
{Michilli}, D., {Seymour}, A., {Hessels}, J.~W.~T., {et~al.} 2018, \nat, 553,
  182, \dodoi{10.1038/nature25149}

\bibitem[{{Nimmo} {et~al.}(2020){Nimmo}, {Hessels}, {Keimpema}, {Archibald},
  {Cordes}, {Karuppusamy}, {Kirsten}, {Li}, {Marcote}, \& {Paragi}}]{Nimmo_20}
{Nimmo}, K., {Hessels}, J.~W.~T., {Keimpema}, A., {et~al.} 2020, arXiv
  e-prints, arXiv:2010.05800.
\newblock \doarXiv{2010.05800}

\bibitem[{{Pearlman} {et~al.}(2020){Pearlman}, {Majid}, {Prince}, {Nimmo},
  {Hessels}, {Naudet}, \& {Kocz}}]{Pearlman+20}
{Pearlman}, A.~B., {Majid}, W.~A., {Prince}, T.~A., {et~al.} 2020, \apjl, 905,
  L27, \dodoi{10.3847/2041-8213/abca31}

\bibitem[{{Petroff} {et~al.}(2019){Petroff}, {Hessels}, \&
  {Lorimer}}]{Petroff+19}
{Petroff}, E., {Hessels}, J.~W.~T., \& {Lorimer}, D.~R. 2019, \aapr, 27, 4,
  \dodoi{10.1007/s00159-019-0116-6}

\bibitem[{{Petroff} {et~al.}(2021){Petroff}, {Hessels}, \&
  {Lorimer}}]{Petroff+21}
---. 2021, arXiv e-prints, arXiv:2107.10113.
\newblock \doarXiv{2107.10113}

\bibitem[{{Philippov} {et~al.}(2019){Philippov}, {Uzdensky}, {Spitkovsky}, \&
  {Cerutti}}]{Philippov+19}
{Philippov}, A., {Uzdensky}, D.~A., {Spitkovsky}, A., \& {Cerutti}, B. 2019,
  \apjl, 876, L6, \dodoi{10.3847/2041-8213/ab1590}

\bibitem[{{Platts} {et~al.}(2019){Platts}, {Weltman}, {Walters}, {Tendulkar},
  {Gordin}, \& {Kandhai}}]{Platts+19}
{Platts}, E., {Weltman}, A., {Walters}, A., {et~al.} 2019, \physrep, 821, 1,
  \dodoi{10.1016/j.physrep.2019.06.003}

\bibitem[{{Pleunis} {et~al.}(2021){Pleunis}, {Good}, {Kaspi},
  {et~al.}}]{Pleunis+21}
{Pleunis}, Z., {Good}, D.~C., {Kaspi}, V.~M., {et~al.} 2021, arXiv e-prints,
  arXiv:2106.04356.
\newblock \doarXiv{2106.04356}

\bibitem[{{Plotnikov} \& {Sironi}(2019)}]{Plotnikov&Sironi19}
{Plotnikov}, I., \& {Sironi}, L. 2019, \mnras, 485, 3816,
  \dodoi{10.1093/mnras/stz640}

\bibitem[{{Rajwade} {et~al.}(2020){Rajwade}, {Mickaliger}, {Stappers},
  {Morello}, {Agarwal}, {Bassa}, {Breton}, {Caleb}, {Karastergiou}, {Keane}, \&
  {Lorimer}}]{Rajwade+20}
{Rajwade}, K.~M., {Mickaliger}, M.~B., {Stappers}, B.~W., {et~al.} 2020,
  \mnras, 495, 3551, \dodoi{10.1093/mnras/staa1237}

\bibitem[{{Scholz} {et~al.}(2016){Scholz}, {Spitler}, {Hessels}, {Chatterjee},
  {et~al.}}]{Scholz+16}
{Scholz}, P., {Spitler}, L.~G., {Hessels}, J.~W.~T., {Chatterjee}, S., {et~al.}
  2016, \apj, 833, 177, \dodoi{10.3847/1538-4357/833/2/177}

\bibitem[{{Shannon} {et~al.}(2018){Shannon}, {Macquart}, {Bannister}, {Ekers},
  {James}, {Os{\l}owski}, {Qiu}, {Sammons}, {Hotan}, {Voronkov}, {Beresford},
  {Brothers}, {Brown}, {Bunton}, {Chippendale}, {Haskins}, {Leach},
  {Marquarding}, {McConnell}, {Pilawa}, {Sadler}, {Troup}, {Tuthill},
  {Whiting}, {Allison}, {Anderson}, {Bell}, {Collier}, {G{\"u}rkan}, {Heald},
  \& {Riseley}}]{Shannon+18}
{Shannon}, R.~M., {Macquart}, J.~P., {Bannister}, K.~W., {et~al.} 2018, \nat,
  562, 386, \dodoi{10.1038/s41586-018-0588-y}

\bibitem[{{Simard} \& {Ravi}(2020)}]{Simard&Ravi20}
{Simard}, D., \& {Ravi}, V. 2020, \apjl, 899, L21,
  \dodoi{10.3847/2041-8213/abaa40}

\bibitem[{{Sironi} {et~al.}(2021){Sironi}, {Plotnikov}, {N{\"a}ttil{\"a}}, \&
  {Beloborodov}}]{Sironi+21}
{Sironi}, L., {Plotnikov}, I., {N{\"a}ttil{\"a}}, J., \& {Beloborodov}, A.~M.
  2021, arXiv e-prints, arXiv:2107.01211.
\newblock \doarXiv{2107.01211}

\bibitem[{{Sobacchi} {et~al.}(2021){Sobacchi}, {Lyubarsky}, {Beloborodov}, \&
  {Sironi}}]{Sobacchi+21}
{Sobacchi}, E., {Lyubarsky}, Y., {Beloborodov}, A.~M., \& {Sironi}, L. 2021,
  \mnras, 500, 272, \dodoi{10.1093/mnras/staa3248}

\bibitem[{{Sridhar} {et~al.}(2021{\natexlab{a}}){Sridhar}, {Metzger},
  {Beniamini}, {Margalit}, {Renzo}, {Sironi}, \& {Kovlakas}}]{Sridhar+21b}
{Sridhar}, N., {Metzger}, B.~D., {Beniamini}, P., {et~al.} 2021{\natexlab{a}},
  \apj, 917, 13, \dodoi{10.3847/1538-4357/ac0140}

\bibitem[{{Sridhar} {et~al.}(2021{\natexlab{b}}){Sridhar}, {Zrake}, {Metzger},
  {Sironi}, \& {Giannios}}]{Sridhar+21}
{Sridhar}, N., {Zrake}, J., {Metzger}, B.~D., {Sironi}, L., \& {Giannios}, D.
  2021{\natexlab{b}}, \mnras, 501, 3184, \dodoi{10.1093/mnras/staa3794}

\bibitem[{{The CHIME/FRB Collaboration} {et~al.}(2021){The CHIME/FRB
  Collaboration}, {:}, {Amiri}, {Andersen}, {Bandura},
  {et~al.}}]{CHIME+21_catalog}
{The CHIME/FRB Collaboration}, {:}, {Amiri}, M., {et~al.} 2021, arXiv e-prints,
  arXiv:2106.04352.
\newblock \doarXiv{2106.04352}

\bibitem[{{The CHIME/FRB Collaboration} {et~al.}(2020){The CHIME/FRB
  Collaboration}, {:}, {Andersen}, {et~al.}}]{CHIME+20}
{The CHIME/FRB Collaboration}, {:}, {Andersen}, B.~C., {et~al.} 2020, arXiv
  e-prints, arXiv:2005.10324.
\newblock \doarXiv{2005.10324}

\bibitem[{{Tuntsov} {et~al.}(2021){Tuntsov}, {Pen}, \& {Walker}}]{Tuntsov+21}
{Tuntsov}, A., {Pen}, U.-L., \& {Walker}, M. 2021, arXiv e-prints,
  arXiv:2107.13549.
\newblock \doarXiv{2107.13549}

\bibitem[{{Wadiasingh} \& {Timokhin}(2019)}]{Wadiasingh2019}
{Wadiasingh}, Z., \& {Timokhin}, A. 2019, \apj, 879, 4,
  \dodoi{10.3847/1538-4357/ab2240}

\end{thebibliography}

\end{document}